\documentclass[epjH]{svjour}
\pdfoutput=1
\usepackage{graphics}
\usepackage{sidecap}
\usepackage{amssymb}
\usepackage{wrapfig}
\usepackage{units}
\usepackage{xcolor}
\usepackage{natbib}
\usepackage{hyperref}
\setcitestyle{aysep}
\setcitestyle{square}

\def\MeV{\ifmmode {\mathrm{\ Me\kern -0.1em V}}\else
                   \textrm{Me\kern -0.1em V}\fi\xspace}%
\def\GeV{\ifmmode {\mathrm{\ Ge\kern -0.1em V}}\else
                   \textrm{Ge\kern -0.1em V}\fi\xspace}%
\def\TeV{\ifmmode {\mathrm{\ Te\kern -0.1em V}}\else
                   \textrm{Te\kern -0.1em V}\fi\xspace}%
\def\PeV{\ifmmode {\mathrm{\ Pe\kern -0.1em V}}\else
                   \textrm{Pe\kern -0.1em V}\fi\xspace}%
\def\EeV{\ifmmode {\mathrm{\ Ee\kern -0.1em V}}\else
                   \textrm{Ee\kern -0.1em V}\fi\xspace}%
\def\eV{\ifmmode {\mathrm{\ e\kern -0.1em V}}\else
                   \textrm{e\kern -0.1em V}\fi\xspace}%
\def\gcm{g\kern 0.1em cm$^{-2}$}

\begin{document}
\title{Extensive Air Showers and Ultra High-Energy Cosmic Rays: A Historical Review}

\author{Karl-Heinz Kampert\inst{1}\fnmsep\thanks{\email{kampert@uni-wuppertal.de}} 
\and Alan A Watson\inst{2}\fnmsep\thanks{\email{a.a.watson@leeds.ac.uk}}}

\institute{Department of Physics, University Wuppertal, Germany\\ \and School of Physics and Astronomy, University of Leeds, UK
}

\abstract{
The discovery of extensive air showers by Rossi, Schmeiser, Bothe, Kolh\"orster and Auger at the end of the 1930s, facilitated by the coincidence technique of Bothe and Rossi, led to fundamental contributions in the field of cosmic ray physics and laid the foundation for high-energy particle physics. Soon after World War II a cosmic ray group at MIT in the USA pioneered detailed investigations of air shower phenomena and their experimental skill laid the foundation for many of the methods and much of the instrumentation used today. Soon interests focussed on the highest energies requiring much larger detectors to be operated.  The first detection of air fluorescence light by Japanese and US groups in the early 1970s marked an important experimental breakthrough towards this end as it allowed huge volumes of atmosphere to be monitored by optical telescopes. Radio observations of air showers, pioneered in the 1960s, are presently experiencing a renaissance and may revolutionise the field again. In the last 7 decades the research has seen many ups but also a few downs. However, the example of the Cygnus X-3 story demonstrated that even non-confirmable observations can have a huge impact by boosting new instrumentation to make discoveries and shape an entire scientific community.
}

\maketitle

\date{\today}

\section{Introduction and General Overview }
\label{sec:gen-overview}

Towards the end of the 1930s it was recognised from studies of the effect of the geomagnetic field on cosmic rays that the energy spectrum of the primary particles, not identified as being proton-dominated until 1941, extended to at least 10~GeV.  The discovery of extensive air showers in 1938, however, radically changed this situation with the highest energy being pushed up by about 5 orders of magnitude, probably the single largest advance to our knowledge of energy scales ever made.  It is now known that the energy spectrum extends to beyond $10^{20}$~eV but it has taken over 60 years to consolidate this picture.  In this section we trace the history of the discovery of extensive air showers, show how advances in experimental and theoretical techniques have led to improved understanding of them, and describe how some of the most recent work with contemporary instruments has provided important data on the energy spectrum, the mass composition and the arrival direction distribution of high-energy cosmic rays.  These results are of astrophysical importance but additionally some aspects of the shower phenomenon promise to give new insights on hadronic physics at energies beyond that reached by the LHC.

The flux of particles falls so rapidly with energy ($\propto E^{-\gamma}$ with $\gamma \sim 2.7$) that around $10^{14}$~eV it becomes impractical to make measurements of high precision directly: the number of events falling on a detector of a size that can be accommodated on a balloon or a space-craft is simply too small.  However at this energy sufficient particles are produced in the atmosphere as secondaries to the incoming primary cosmic rays for some to reach mountain altitudes and, as the energy of the primary increases, even sea level.  The transverse momentum acquired by secondary particles at production and the scattering which the shower electrons, in particular, undergo through interactions with the material of the atmosphere are such that the secondaries are spread over significant areas at the observational level.  The phenomenon of the nearly-simultaneous arrival of many particles over a large area is called an Extensive Air Shower (EAS): at $10^{15}$~eV around $10^6$ particles cover approximately $10^4$~m$^2$ while at $10^{20}$~eV some $10^{11}$~particles are spread over about 10~km$^2$.   
It was quickly recognised that the phenomenon of the air shower offered the possibility of answering four major questions:

\begin{enumerate}
\item{{\bf What particle physics can be learned from understanding air shower evolution?}}

A detailed understanding of how an air shower develops is crucial to obtaining an estimate of the primary energy and to learning anything about the mass spectrum of the primary particles.  It is worth recalling that when the shower phenomenon was first observed that, in addition to the proton, neutron, electron and positron, only the muon was known, so that a realistic understanding of shower development had to wait until the discovery of the charged pion and its decay chain in 1947 and of the neutral pion in 1950.  Indeed, much early thinking was based on the hypothesis that showers were initiated by electrons and/or photons.  Once it was recognised that the initiating particle was almost always a proton or a nucleus, the first steps in understanding the nuclear cascade focussed on such matters as whether a proton would lose all or only part of its energy in a nuclear collision and how many pions were radiated in such a collision.  A combination of observations in air showers, made using Geiger counters and cloud chambers, of data from studies in nuclear emulsions and of early accelerator information was used to inform the debate.  The issues of inelasticity (what fraction of the energy is lost by an incoming nucleon to pion production) and the multiplicity (the number of pions produced) are parameters which are still uncertain at most of the energies of interest.

\item{{\bf What can be inferred from the arrival direction distributions of the high-energy particles?}}

From the earliest years of discovery of cosmic rays there have been searches for directional anisotropies.  Hess himself, from a balloon flight made during a solar eclipse in April 1912, i.e.\ before his discovery flight in August of the same year, deduced that the Sun was not a major source \citep{Hess:1912ui}.
There are a few predictions of the level of anisotropy that might be expected.  While there have always been speculations as to the sources, the fact that the primary particles are charged and therefore are deflected in the poorly-known galactic and intergalactic magnetic fields makes it difficult to identify them.  One firm prediction was made very early on by Compton and Getting in 1935
\citep{compton35} that cosmic rays should show an anisotropy because of the motion of the earth within the galaxy.  Eventually it was realised that this idea would be testable only with cosmic rays undeflected by the solar wind (discovered much later) so measuring the Compton-Getting effect became a target for air shower experiments.  However, as the velocity of the earth is only about 200~km\,s$^{-1}$, the effect is $\sim 0.1$\,\% and it has taken around 70 years for a convincing demonstration of its discovery.  The search for point sources has been largely unsuccessful but one of the motivations for searching for rarer and rarer particles of higher and higher energy has been the expectation that anisotropy would eventually be found.

\item{{\bf What is the energy spectrum of the primary cosmic rays?}}

A power law distribution of cosmic rays was first described by E Fermi in 1949 \citep{Fermi-49} but until 1966 there were no predictions as to the power law index or to further structures in the energy spectrum. Observations in 1959 had indicated a steepening 
at around $3 \cdot 10^{15}$~eV (the ``knee''),
while in 1963 it was claimed from observations made with the first large shower array that the spectrum flattens just above $10^{18}$~eV.  However not only were there no predictions of these features, interpretation of them remains controversial.  By contrast the discovery of the 2.7~K cosmic background radiation in 1965 led, a year later, to the firm statement that if cosmic rays of energy above $\sim 4 \cdot 10^{19}$~eV exist they can come only from nearby sources.  It took about 40 years to establish that there is indeed a steepening in the cosmic ray spectrum at about this energy but whether this is a cosmological effect or a consequence of a limit to which sources can accelerate particles is unclear: $4 \cdot 10^{19}$~eV is within a factor of $\sim 5$ of the highest energy event ever recorded.

\item{{\bf What is the mass composition of the primary cosmic rays?}}

One of the major tasks of the air shower physicist is to find the mass of the primary particles.  This has proved extraordinarily difficult as even if the energy of the primary that produces an event is known, the uncertainties in the hadronic physics make it hard to separate protons from iron.  Data from the LHC will surely help but above $10^{17}$~eV one has reached a regime where the centre-of-mass (cms) energies in the collisions are above what is accessible to man-made machines.  Indeed it may be that in the coming decades the highest-energy cosmic rays provide a test bed for theories of hadronic interactions, mirroring the fact that cosmic ray physics was the place where particle physics was born in the 1930s.

\end{enumerate}

In what follows we have chosen to emphasise the progress made since the 1940s towards answering these four questions through an examination of the development of different techniques, both experimental and analytical, introduced in the last 70 years.  While new techniques have enabled air showers to be studied more effectively, it is remarkable how the essentials of what one seeks to measure were recognised by the pioneers in the 1940s and 1950s.  Increasingly sophisticated equipment, operated on increasingly larger scales has been developed and has led to some answers to the key questions although many issues remain uncertain.  

Galbraith \citep{Galbraith-58} and Cranshaw \citep{Cranshaw-63} have written books in which details of early work, up to the end of the 1950s, are discussed in more detail than is possible below while in Hillas's classic book on Cosmic Rays \citep{Hillas-72} there is an excellent discussion of some of the earliest papers in a context which includes fundamental ideas of cosmic rays physics, including shower physics.

We now move on by reviewing the history of the discovery of the air shower phenomenon.

\section{The Discovery of Extensive Air Showers}
\label{EAS}

\begin{figure}[t]
\centerline{\resizebox{0.45\textwidth}{!}{\includegraphics{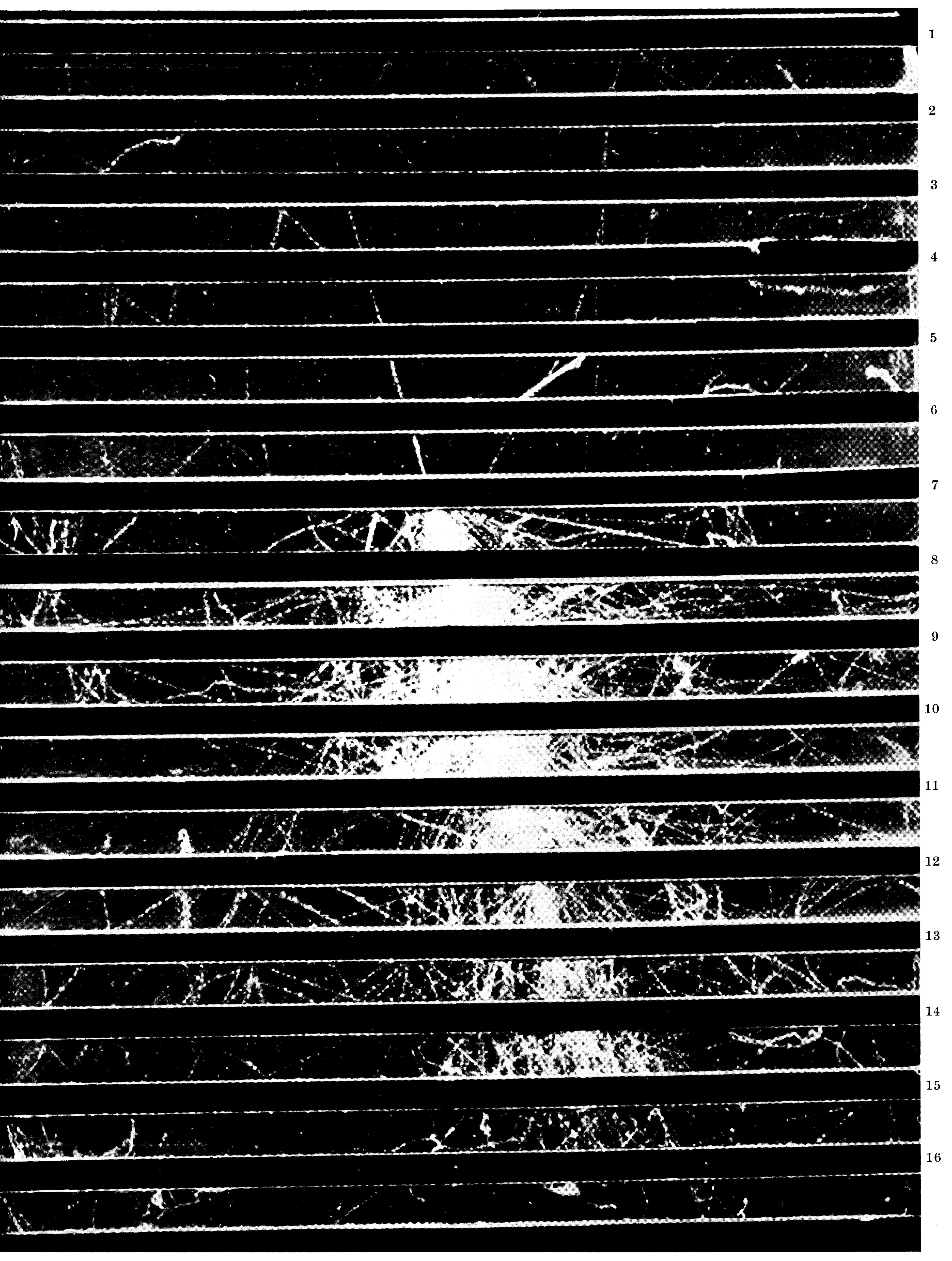}}}
\caption{
Image of a particle cascade, or shower, as seen in a cloud chamber at 3027 m altitude. The primary particle is estimated to be a proton of about 10 GeV. The first interaction will most probably have been in one of the lead plates.  Neutral pions feed the cascade which multiplies in the lead.  Charged pions make similar interactions to protons, or decay into muons.  The cross-sectional area of the cloud chamber is $0.5 \times 0.3$~m$^2$ and the lead absorbers have a thickness of 13~mm each \citep{Fretter-49}.}.
\label{fig:cloud-chamber}
\end{figure}

A technical developments of crucial importance for the study of cosmic rays was the invention of the coincidence technique by Walther Bothe in the late 1920s \citep{Bothe:1929vl} for which he was awarded the Nobel Prize in 1954. Coupling the coincidence technique to the newly developed fast responding Geiger-M\"uller counters \citep{GM-28} had already allowed to verification that Compton scattering produces a recoil electron simultaneously with the scattered $\gamma$-ray. Bothe's coincidence circuit reached a resolving time for singly charged particles of 1.4~ms but was limited to only twofold coincidences. Only few months later, Rossi described a coincidence circuit which was conceptually different from Bothe's as it could accommodate many channels \citep{Rossi:1930uw}. He also pushed the resolving time down to 0.4\,ms. This, together with the strong reduction of accidentals in triple coincidences, allowed for the detection of rare cosmic events. In the mid-1930s the coincidence method has also been used to trigger a cloud chamber inside a magnetic field. Instead of using the usual method of random expansion of the chamber, as had to be performed by Dimitry Skobeltzyn for his discovery of multiple production of fast $\beta$-particles in single interaction processes \citep{Skobeltzyn:1927wa,Skobeltzyn:1929wb},  Blackett and Occhialini \citep{Blackett-Occhialini-1932} placed Geiger-M\"uller counters above and below a vertical cloud chamber, so that charged particles passing through the two counters would also pass through the chamber,
triggering its expansion. This technique allowed the observation of apparently simultaneous production of numerous electrons and positrons much more effectively (cf.\ Fig. \ref{fig:cloud-chamber}).
Blackett in his Nobel lecture of 1948 recalled ``that the development of the counter-controlled cloud chamber method, not only attained the original objective of achieving much economy in both time and film, but proved to have the unexpected advantage of greatly enhancing the number of associated rays photographed'' \citep{Blackett-48}. In retrospect, this experiment marked the birth of ``rare event triggering'', which became a key tool for making progress in nuclear and particle physics experiments.

\begin{figure}[t]
\centerline{\resizebox{0.9\textwidth}{!}{\includegraphics{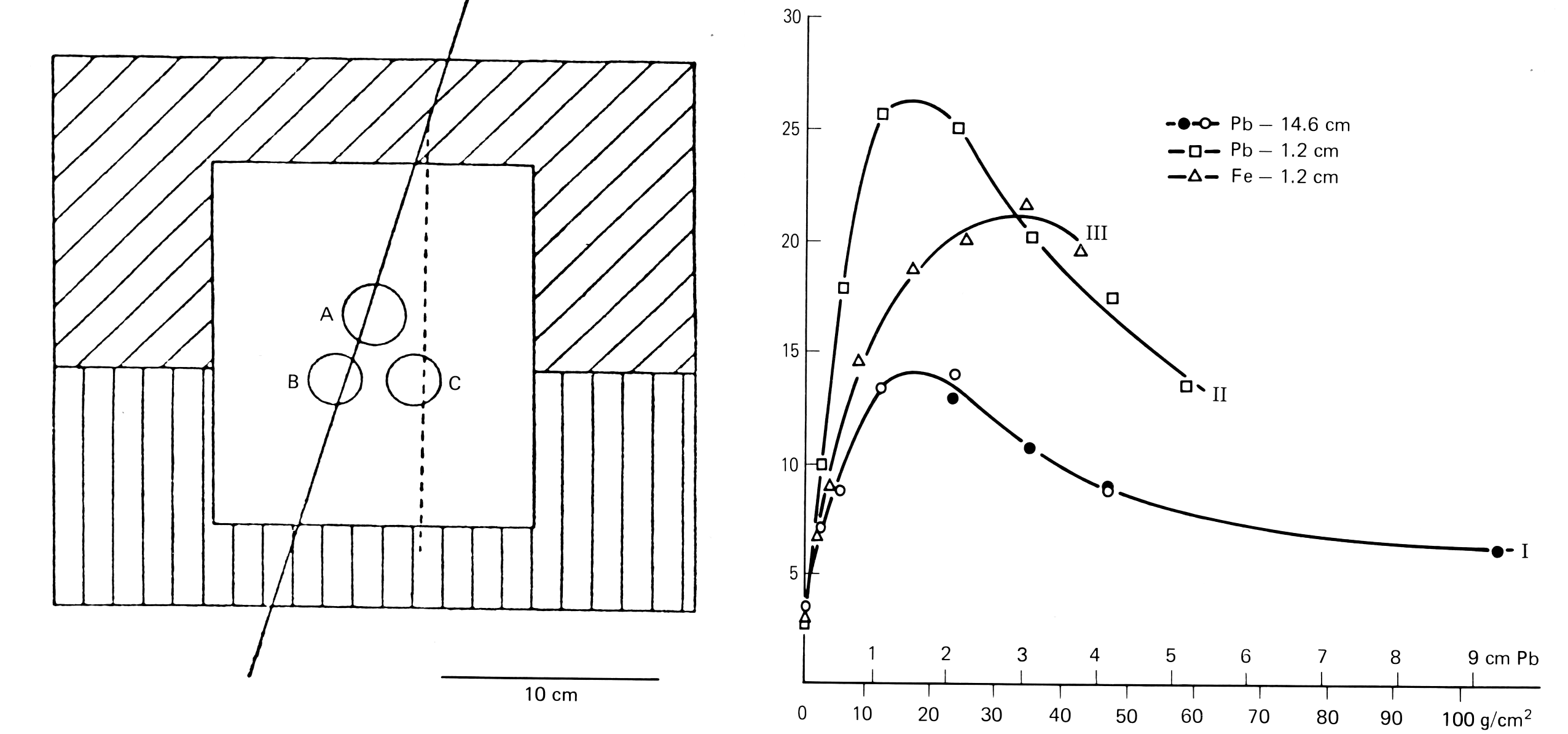}}}
\caption{Rossi's transition curve: The experiment in which the abundant production of secondary radiation by cosmic rays was discovered. Coincidences between Geiger-M\"uller counters, arranged as on the left, are produced by groups secondary particles generated by cosmic rays in the lead shield above the counters. The curves labeled I-III refer to p and Fe absorbers of different thicknesses placed above the counters \citep{Rossi:1933vw}. \label{fig:rossi-trans-curve}}
\end{figure}

The development of the coincidence approach was crucial also for the discovery and study of extensive air showers. In 1933 Rossi made a key observation which was hard to accept for the scientific community and which, as Rossi recalled later \citep[page 71]{Rossi-85}, even ``raised doubts about the legitimacy of the coincidence method''. When extending previous measurements by Bothe and Kolh\"orster about the absorption of cosmic rays to a maximum of 101 cm of lead, he concluded that 50\,\% of the rays could penetrate a metre of lead for which the maximum particle energy exceeded $1.4 \cdot 10^{10}$~eV based on energy-loss estimates by Heisenberg \citep{Heisenberg:1932ur}. All this gave support to the corpuscular nature of cosmic rays in agreement with conclusions by Bothe and Kolh\"orster. The key result of this paper later became known as {\em ``Rossi's transition curve''}. Rossi observed a rapid {\em increase} of triple coincidences in a triangular arrangement of Geiger counters (c.f.\ Fig.~\ref{fig:rossi-trans-curve}) when some centimetres of lead was placed above \citep{Rossi:1933vw}. Only with with further increasing absorber thickness did the coincidence rate start to decline. Rossi correctly concluded that soft secondary particles were produced by cosmic particles entering the material. These secondary particles then suffer increasing absorption with increasing total thickness of the absorber. It is interesting to note that the same basic observation was made a year later by Regener and Pfotzer \citep{Regener:1935uv} when studying the vertical intensity of cosmic rays in the stratosphere up to a height of 28~km by recording the rate of threefold coincidences. Flying and operating sensitive instruments in the stratosphere was a remarkable experimental achievement in itself which became possible because of Regener's long term experience in flying balloon-borne instruments for atmospheric studies and because of his tedious work in patching hundreds of tiny pinholes in the rubber balloons to prevent untimely bursting of the balloons in the upper atmosphere. All this work paid off by observing an unexpected clear maximum in the coincidence rate at a pressure of 100~mm of mercury (about 14~km above sea level). This became known as the ``Pfotzer Maximum''. Regener correctly interpreted \citep{Regener:1938tm} the maximum as being due to the multiplication of electrons in the atmosphere -- which he called {\em ``Schauer''} -- such as had been suggested by Bhabha and Heitler \citep{Bhabha-37}. However, neither Rossi nor Regener seem to have recognized that the same physical mechanism was behind their observations.

\begin{figure}[t]
\centerline{\resizebox{0.6\textwidth}{!}{\includegraphics{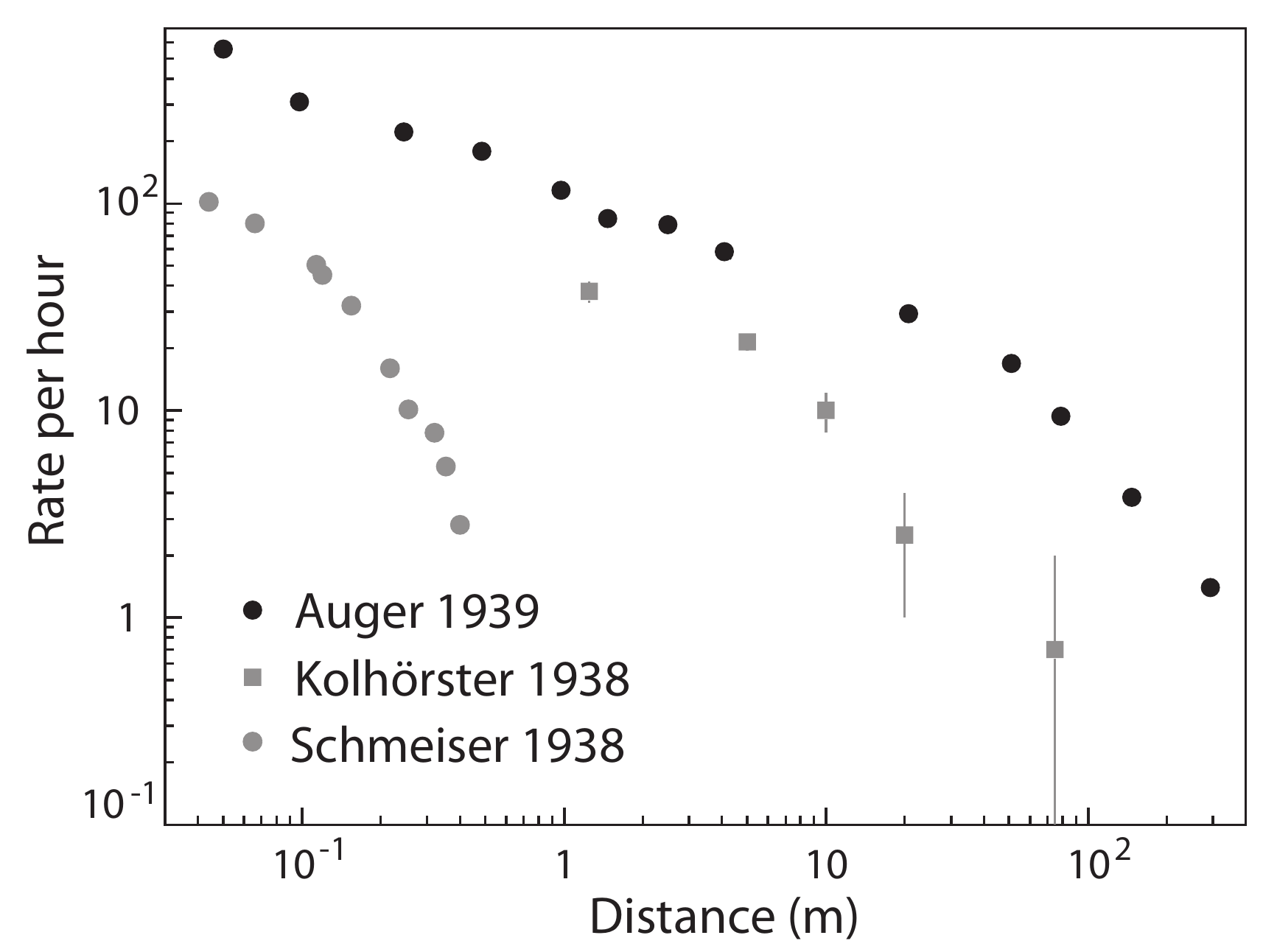}}}
\caption{The discovery of extensive air showers: Decoherence curves measured with Geiger counters separated up to 300~m distance. Data of \citep{Schmeiser:1938us} and \citep{Kolhorster:1938vm} were measured at sea level with counters of 91 cm$^2$ and 430 cm$^2$ effective area, respectively, while data of \citep{Auger:1939wp} were measured with counters of 200 cm$^2$ at the Jungfraujoch at 3450 m. \label{fig:coincidence-rate}}
\end{figure}

Schmeiser and Bothe pointed out that Rossi's transition curve implied the occurrence of showers in air -- which they named {\em ``Luftschauer''} -- and showed that particles in air showers had separations up to 40~cm \citep{Schmeiser:1938us}.  Independently, Kolh\"orster et al.\ \citep{Kolhorster:1938vm} reported data on the rate at which coincidences between a pair of Geiger counters fell as a function of separation. The results of these pioneering measurements are shown in Fig.~\ref{fig:coincidence-rate}.  It is clear, however, that Rossi had made the same discovery some years earlier.  In 1934, he made observations in Eritrea that suggested to him that there was a correlated arrival of particles at widely-separated detectors.  In his publication \citep{Rossi:1934vt} he gave the phenomenon the name {\em ``sciami''}. He was not able to follow up this work before he had to leave Italy and it seems to have been unknown to either Bothe or Kolh\"orster.  

Despite the work of Rossi and the two German groups, credit for the discovery of extensive air showers has usually been given to Auger and his collaborators for what seems to have been a serendipitous observation \citep{Auger:1939wp} depending strongly on the electronic developments by Roland Maze who improved the resolving time of coincidence circuits to 5~$\mu$s \citep{Maze:1938cu}.  Auger, Maze and Robley  found that the chance rate between two counters separated by some distance greatly exceeded the chance rate expected from the resolving time of the new circuit.  For a while the phenomenon was known as ``Auger showers'' \citep[page 214]{Auger-85}.
In their measurements performed at the Jungfraujoch in the Swiss Alps they were able to separate their detectors by up to 300~m. The decoherence curves are shown again in Fig.~\ref{fig:coincidence-rate}. Differences in the coincidence rates between the three groups of authors can be understood both by the different effective areas of the Geiger counters and by the different altitudes at which the measurements were performed. In view of the sequence of air shower observations, the important achievement of Auger and his group, what distinguishes their work from that of Rossi, Schmeiser \& Bothe, and Kolh\"orster appears not so much in separating their detectors by up to 300\,m, but in estimating the primary energy to be around $10^{15}$~eV. This estimate was based on the number of particles in the showers, assuming that each particle carried, on average, the critical energy. A factor of 10 was added to account for the energy lost in the atmosphere.
A similar conclusion came from using the work of Bhabha and Heitler, based on the ideas of quantum electrodynamics (QED).  It is worth quoting the final remarks of Auger from his paper presented at the 1939 Symposium held in Chicago \citep{auger-39}:

\begin{quote}
{\it One of the consequences of the extension of the energy spectrum of cosmic rays up to $10^{15}$~eV is that it is actually impossible to imaging a single process able to give a particle such an energy.  It seems much more likely that the charged particles which constitute the primary cosmic radiation acquire their energy along electric fields of very great extension.}
\end{quote}

The identification of mechanisms to accelerate particles to energies as great as $10^{20}$~eV that have now been observed remains a great challenge, though the mechanism suggested by Auger now seems unlikely as electric fields over great extensions are unavailable because of the conductivity of the interstellar plasma.

Several groups, including Auger's, verified the inferences drawn from the Geiger-counter observations using cloud chambers.  Of some interest to the evolution of shower activities in the UK is the work that was done in Manchester where
A Lovell and J G Wilson \citep{Lovell-39}, using cloud chambers separated by 5.5 and 19~m, observed parallel tracks in the cloud chambers, one of which was triggered by an array of Geiger counters.  It appears that this study was encouraged by Auger, who was visiting Blackett's laboratory in Manchester prior to his departure for Chicago where he spent the war years \citep{Lovell-08}.
Rossi was in Blackett's laboratory for six months from late 1938 and developed the anti-coincidence technique there.
Blackett's interest in extensive air showers was stimulated by Auger and this led to his close involvement in shower activities in Britain post-WW\,II.

While in Chicago, Auger carried out several experiments in which he launched detectors carried by several balloons to study the way in which air showers behaved at altitude.
He also attended the Chicago conference on cosmic rays held in 1939 and made a major report on his work \citep{auger-39}.

That Bothe, Kolh\"orster and Auger seem to have been unaware of Rossi's work perhaps reflects the fact that the research was done at a time when scientists wrote most commonly in their native languages: in addition, there was no preprint system such as operated in the post-war period and, of course, there was no arXiv.  Information about new results was sometimes exchanged by correspondence between senior scientists or during face-to-face meetings.  The prominence given to Auger's work probably arises from his stay with Blackett in Manchester and he was able to take advantage of his time in Chicago in the early 1940s relatively unhindered by war work. Presumably Bothe, who attended the Chicago meeting, and Kolh\"orster had little chance for cosmic ray work after 1939.  Rossi left Manchester for Chicago and before joining the Manhattan project, his work was focussed largely on the problem of muon decay.

Only a few years after the discovery of extensive air showers, Skobeltzyn, George Zatsepin, and V Miller \citep{Skobeltzyn:1947tv} at the Pamir mountains at an altitude of 3860\,m above sea level pushed  measurements of coincidences out to distances of 1000~m. To suppress random coincidences which would occur between single distant Geiger counters, they were the first to apply so-called double-coincidences, meaning that coincidences were first formed within trays of local Geiger counters, before a coincidence was formed between the distant trays.


\section{Basic Ideas about Extensive Air Showers}
\label{sec:BasicIdeas}

Work by Auger and his colleagues using cloud chambers triggered by arrays of Geiger counters allowed features of air showers to be understood relatively quickly.  By the late 1930s it was known that air showers contained hadronic particles, muons and electrons and major advances in understanding took place in the late 1940s and early 1950s after the existence of two charged and one neutral pion was established and it was recognised that muons were secondary to charged pions.  The development of an air shower can be understood by studying Fig.\,\ref{fig:cloud-chamber} which we will reference on occasion.  In the figure a cloud chamber picture of a shower created in lead plates by a cosmic ray proton of about 10 GeV is shown \citep{Fretter-49}.  The features visible in this photograph, except for scale, are extremely similar to those present when a high-energy particle enters the earth's atmosphere and creates a shower.  

Each lead plate (the dark bands running horizontally across the picture) is about two radiation-lengths\footnote{The radiation-length is an appropriate scale length for describing high-energy electromagnetic cascades. It is both the mean distance over which a high-energy electron loses all but $1/e$ of its energy by bremsstrahlung, and $7/9$ of the mean free path for pair production by a high-energy photon.} thick and the cross-sectional area of the cloud chamber is  $0.5 \times 0.3$~m$^2$.  The gas in the chamber was argon, effectively at atmospheric pressure, and thus most of the shower development happens within the lead plates.  Little development of the cascade takes place in the gas but the level of condensation gives a snapshot of how the particle number increases and decreases as the shower progresses through more and more lead.  All of the important features of shower development, such as the rise and fall of the particle numbers (which in an EAS experiment will be called {\em ``shower size''}), and the lateral spreading of the shower, are evident, as are some muons that penetrate more deeply into the chamber than most of the electrons.  

The incoming particle can be identified as a proton of about 10\,GeV with some confidence.
Had it interacted near sea-level in air then the extent of the shower lateral spread of the shower would have been around 50~m.
The level of ionisation excludes a heavier nucleus and the traversal of the particle through six lead plates (about 88.5\,\gcm \,or 13.9 radiation lengths) strongly excludes the possibility that the incoming particle is an electron.  To have the point of interaction, presumably with a lead nucleus, in the 7$^{\rm th}$ plate is very reasonable, as it is now known that the $p$-air cross-section is around 250~mb (equivalent to 80\,\gcm) at the energy estimated for the proton, and that the interaction length in lead is about 194\,\gcm.
The interaction of a primary proton with an absorber nucleus $A$ can be represented as
\begin{equation}
p + A \to p + X + \pi^{\pm,0} + K^{\pm,0} ...
\end{equation}
with $X$ representing the fragmented nucleus. Of course, charge and all the other quantum numbers must be conserved. The proton in the exit channel is called the ``leading particle'', typically carrying about 50\,\% (which is called ``inelasticity'') of the incoming energy. Elementary particles other than pions will be created but it was demonstrated experimentally, from studies of muons in air showers, that the cross-section for the creation of kaons and hyperons in such collisions are down by a factor of around 10, at least at energies up to $\sim 10^{15}$~eV \citep[Sec.\,4]{Greisen-60}.

The problem of identifying the nature and determining the energy of the particle that initiated this shower, if there were data available from only one layer of gas corresponding to the information available from a shower array at a single atmospheric depth, can be appreciated from Fig.\,\ref{fig:cloud-chamber}.  But until the 1980s, when a technique was developed that allowed the build-up of the air shower to be studied on an event-by-event basis, as is seen in the figure, this was the challenge faced by all air shower experimenters.   Assumptions had to be made as to where the particle had its first interaction and what are the features of the hadronic interactions.  Key parameters such as the cross-sections for the interaction of protons (and heavier nuclei) with nuclei, pion-nucleus cross-sections, the fraction of energy radiated as pions in each collision and the number of particles produced are needed.  By contrast determination of the direction of the incoming primary is a relatively straight-forward exercise compared with those of finding the mass and the energy.

\begin{figure}[t]
\centerline{\resizebox{0.85\textwidth}{!}
{\includegraphics{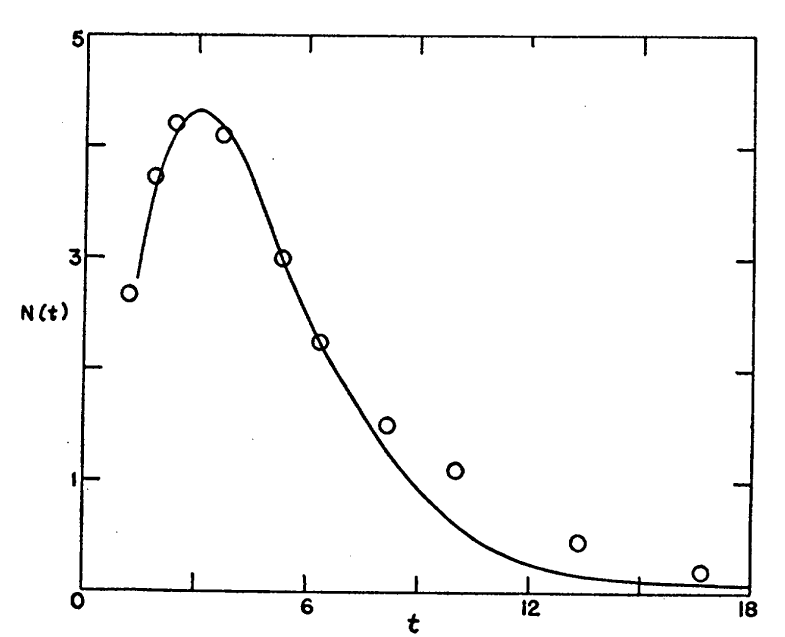}\includegraphics{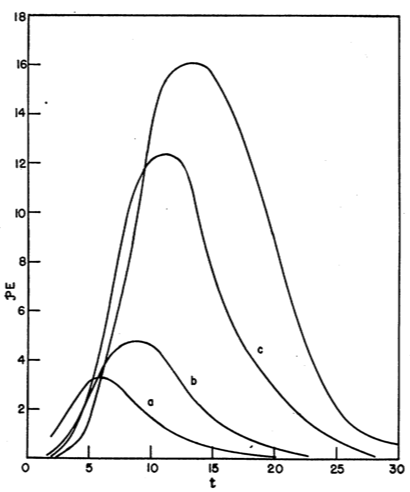}}}
\caption{
Left: Total number of electrons $N(t)$ against $t=X/X_0$ ($X_0$: radiation length), calculated by Carlson and Oppenheimer for 2.5~GeV electrons in air and compared to experimental results (circles) of Pfotzer \citep{Pfotzer:1936vd}. Right: Estimated number of electrons with $E_e>50$\,MeV in Pb for $E_0 = 2.7$, 20, 150, and 1110\,GeV \citep{Carlson:1937uj}.}
\label{fig:carlson-oppenheimer}
\end{figure}

The basic key processes of cascade multiplication occurring in EAS were laid out by Bethe and Heitler based on QED \citep{Bethe:1934tv} and were formulated in terms of pair-production and bremsstrahlung processes by Bhabha and Heitler \citep{Bhabha-37}. Carlson and Oppenheimer \citep{Carlson:1937uj} finally completed the theory by accounting also for energy losses of electrons by ionisation and for practical calculations they pioneered the use of diffusion equations. Moreover, they demonstrated quantitative agreement of their calculations with the aforementioned experimental results by Regener and Pfotzer (Pfotzer-Maximum) (l.h.s. of Fig.~\ref{fig:carlson-oppenheimer}), and pointed out the importance of fluctuations of the shower maximum, and noted that a more penetrating burst like component, as suggested by Heisenberg \citep{Heisenberg:1936vf} based on measurements by Hoffmann\footnote{In 1927 Hoffmann had discovered a phenomenon which became known as ``Hoffmann's bursts'' (Hoffmannsche St\"o{\ss}e) \citep{Hoffmann-1930}. In measurements of ionisation currents in an ionization chamber he found occasional discontinuities of strong currents which were interpreted as nuclear explosions.} was needed to allow electrons to penetrate the atmosphere to a thickness of 30 radiation lengths (r.h.s. of Fig.~\ref{fig:carlson-oppenheimer}). 
This paper presented the simple concept of electromagnetic cascades, as is still found in any textbook and in introductory exercises about high energy particle interactions in matter.
Even though it does not capture accurately all details of electromagnetic showers, it accounts for its most important features: the total number of electrons, positrons, and photons at the shower maximum is simply proportional to the primary energy $E_0$ and the depth of maximum shower development is logarithmically proportional to $E_0$.

Nowadays, particle showers in the atmosphere are simulated on powerful computers by using sophisticated Monte Carlo codes, allowing many more details of interaction features to be added and the cascade model by Bethe and Heitler and Carlson and Oppenheimer is applicable only to primary electrons or photons so that diffusion calculations have largely lost their importance. However, because of its pedagogic advantages the concepts are still used and have been generalized to hadronic primaries, see e.g.\ \citep{matthews-05}.

\section{The Geiger-Counter Era}
\label{Geiger-Counters}

Electrostatic photomultipliers (PMTs) were invented in the late 1930s and became available for cosmic ray studies in the 1950s when studies of Cherenkov-light detection and work with liquid and plastic scintillators started.  Nonetheless significant progress in the understanding of showers was made using arrays of Geiger counters at mountain altitudes in the USSR and at sea-level there and in the UK.  A major limitation of early studies was that the direction of the incoming shower was rarely known and the absorption of showers in atmosphere had to provide a wide-aperture collimator.  In the early years the scale of the experiments, with arrays of a few 10s of metres in diameter, meant that showers from primaries of $10^{14}$ to $10^{16}$~eV were the focus although there was always a drive to find the limiting energy that Nature reached.

In the 1950s an array of Geiger counters that eventually covered $\sim 0.6$~km$^2$ was developed by T Cranshaw and W Galbraith \citep{Cranshaw-54,Cranshaw-57} on the disused airfield at Culham near sea-level, the site of UK Atomic Energy Establishment.  Geiger counters were located at 91 stations on a triangular grid with a spacing of 99~m.  Punched cards were an innovation and were used to detect the pattern of stations fired and the time of the coincidences.
The lack of directional information, other than that provided by atmospheric collimation, was a serious handicap but an important study of the anisotropy of cosmic rays up to energies $\sim 10^{17}$~eV was made with the first stage of the array. 
The level of anisotropy at $10^{17}$~eV was less than 10\,\%, thus ruling out an idea advocated by Richtmyer and Teller \citep{Richtmyer-49} and Alfv\`en \citep{Alfven-49} that cosmic rays were produced by the Sun and became isotropised while stored in a local magnetic field of $\sim 10^{-5}$~G. 


Investigations of air showers in the USSR were initiated by  Skobeltzyn who encouraged Zatsepin of the Lebedev Institute to develop a program in the last years of WW\,II.  Skobeltzyn is usually credited with the first observation of what are now recognised as cosmic rays when in 1927 he observed tracks of particles of several tens of MeV in his cloud chamber during a study of the Compton effect.
The first Russian activity was carried out in the Pamirs (3860~m)
and was the start of a major effort on shower work at mountain stations by Soviet scientists which continued for many decades, latterly at a well-serviced installation at Tien Shan (3340~m)
near Almata. The leaders of this work, in addition to Zatsepin, were N A Dobrotin, S I Nikolsky and S A Slavatinsky.  There was also a major effort in Moscow, headed first by S N Vernov and later by G B Khristiansen.  Until the start of construction of the Yakutsk array in the late 1960s, the Soviet program was largely focussed on studying primary particles of less than $10^{17}$~eV.  Complementary programs were undertaken at Mt Aragatz (3250~m) in Armenia and at Mt Elbrus (4000~m) in Georgia.

A strong motivation of the Russian work was to seek a better understanding of the characteristics of showers (energy flow, fraction of muons and nuclear-active particles).  Complex systems of Geiger counters and ionisation chambers, supplemented with cloud chambers, were employed.  Large hodoscopic systems of up to 2000 counters were built up with appropriate monitoring equipment, requiring teams of 20 to 30 people, a distinctive innovation at a time when a typical group comprised only a few people.  The Russian workers introduced the technique of the ionisation calorimeter in which layers of ionisation chambers were interleaved between lead thus extending the capabilities of the cloud chamber method of measuring the energy of an incoming hadronic particle (c.f.\ Fig.\,\ref{fig:cloud-chamber}).  The aim of constructing these installations was to make measurements of the energy spectrum of `nuclear active particles'\footnote{Russian authors at this time used the term ``nuclear-active'' as a synonym for ``hadronic''} in the cores of extensive air showers.  A key result of the early work in the Pamirs was that pions were the main products of multiple-production processes and a further important conclusion was the development of the `leading particle' concept, promoted particularly by Zatsepin, in which it was argued that a particle came out of a nuclear interaction carrying a large fraction of the energy ($\sim 0.5$) that it had taken into it. The work was also instrumental in providing support for the fact that most showers are not initiated by photons or electrons and also provides an excellent example of how shower studies gave crucial information about hadronic physics at very high energies with $10^{15}$~eV being about the median energy studied.

The study of nuclear-active particles in showers was transferred to Tien Shan in the late 1950s and under the direction of Nikolsky a modernised installation including two ionisation calorimeters (one for studying muons), a large number of scintillators and air-Cherenkov detectors was constructed in the 1960s \citep{Erlykin-65}. 
Some rare observations of deep penetration of showers in the calorimeters were interpreted in terms of unusual features of nuclear reactions, e.g.\ the passive baryon and the long flying component \citep{Aseykin-75, Nikolsky-79}. Reports about these phenomena disappeared with time and it seems to have been an experimental effect and underestimated strengths of fluctuations.

Outside of the USSR studies of the high-energy hadronic component of showers took place in Japan and at Chacaltaya but without the benefit of the massive installations of the Pamirs and Tien Shan.

Parallel activities to study showers at sea-level were developed at Moscow State University under the leadership of Vernov and Khristiansen.  An early configuration with groups of hodoscoped Geiger counters deployed over an area of 800\,m$^2$ was used for a measurement of showers at a primary energy lower than that accessible to the groups working at
MIT\footnote{Massachusetts Institute of Technology} or Cornell (see Sec.\ \ref{sec:MIT} and \ref{sec:Cornell}).
The most important output from this period of Moscow work was the discovery of a feature in the {\em ``size spectrum''} of showers\footnote{The ''shower size spectrum'' or just ''size spectrum'' is a common notion used for the distribution of the shower size, i.e. of the total number of particles that reached ground. The shower size, $N$, is obtained by fitting the lateral distribution $\rho(r)$ of shower particles at ground and evaluating the integral $N=2\pi \int_0^\infty r\rho(r)dr$.}.

In the late 1950s it was recognised that the lateral distribution of showers changed very little with the shower size or with the zenith angle at which the shower impacted.  This is because for the primary energies generally studied ($\sim 10^{15}$ to $10^{16}$~eV) particles detected at the observation level mainly originate from a cascade initiated relative close (a few kilometres) from the ground.  Thus it was possible to argue that a measurement of the density at a known distance from the densest region or core of the shower led directly to a measurement of the total number of particles in the shower.  Studies of the 'density spectrum', measured with a single detector, 
had suggested that there was a steepening above $\sim 1000$~particles per m$^2$.  In an attempt to understand this phenomenon, Kulikov and Khristiansen used their array of Geiger counters and showed that the size spectrum of particles steepens at $N \sim 8 \cdot 10^5$ \citep{Kulikov-59}.  By modern standards, the effect was not very strongly established, in part because it relied on a combination of the Moscow data with data from another experiment \citep{Eidus-52} that had covered a higher size regime to claim the effect, and indeed the authors themselves did not regard the irregularity observed as totally established because of insufficient statistical accuracy.  However this measurement -- the first observation of a structure in the primary cosmic ray energy spectrum, the so-called {\em ``knee''} -- had considerable impact.  It was verified with high precision relatively quickly by a number of groups \citep{Fukui:1960tl,Kameda-60,Allan:1962vk,Kulikov-65}.
%
%
Estimating the energy of the knee from the track-integral method (see Sec.\ \ref{sec:Cornell}), Kulikov and Khristiansen had argued that the break may be caused by diffusion of cosmic rays out of the galaxy, so that cosmic rays at $E>10^{16}$\, eV may have an metagalactic origin. Thus, an astrophysical feature in the cosmic ray spectrum may have been discovered. This started a long running debate, picked up by Peters \citep{peters61} who
proposed that what was being seen reflected a similar feature in the primary spectrum of cosmic rays induced either by a limitation of the acceleration processes or by a leakage of particles from the galaxy.  These are both rigidity effects and depend on the energy per nucleon of the particle so that when protons can no longer be accelerated or leak from the galaxy, the flux of cosmic rays falls and the spectrum steepens.  There were competing claims that this feature was due to a characteristic of nuclear interactions with a change occurring near $10^{15}$~eV.  The debate was not to be settled for a further 45 years until precise data from KASCADE (see section \ref{sec:KASCADE}) became available.  Surprisingly, this distinctive feature of the size spectrum was not discussed in reviews such as that of Greisen \citep{Greisen-60} or in the final discussions of the energy spectrum extracted from the Chacaltaya work discussed later \citep{Bradt-66,LaPointe-68}.

The Moscow University array was continuously enhanced and was one of only three arrays to include a magnetic spectrometer.  Radio emission was also studied there and advantage was taken of a tunnel in the Moscow Metro to study muons above 10~GeV.  The main aim of the work, particularly that led by Khristiansen, was to understand the mass composition, through studies of $N_\mu$ vs.\ $N_e$ and through the steepness of the lateral distribution.
Also for this purpose, the work at Moscow was complemented by very detailed Monte Carlo studies driven by Khristiansen.

When Rossi moved to MIT following his work on the Manhattan project, one strand of activity that he developed was the study of high-energy cosmic rays.  He targeted the problem of determining the energy of the particle initiating each event and of finding its arrival direction.  The plan was to measure the energy by determining the shower size and comparing it with predictions from theoretical models.  He was particularly interested in moving from studies of the average properties of events, the focus hitherto, and guided Robert Williams to the study of individual showers with an array of four fast ionisation-chambers which he used to sample the density of the signals across the shower.  This was an advance over the Geiger counter as the number of particles could be determined directly rather than in a statistical manner.
Williams \citep{Williams-48} made measurements at 3500~m 
(Doolittle Ranch, near Echo Lake) and at 4300~m
(Mt Evans),
using the 4 ionisation chambers arranged in a star-shaped geometry with a central detector and the other three at 7\,m distance.  The time resolution of the ionisation chambers was about 1 $\mu$s, too long to make the determination of directions feasible, but at the lower altitude Williams took advantage of a cloud chamber operated by Wayne Hazen to obtain the arrival direction of a small fraction of the events, finding that $\sim 80$\,\% of them had zenith angles less than $20^\circ$.  He was thus able to make his analysis assuming that the showers were vertical.  By preparing charts of signal size as a function of distance using the shower theory then available, he could estimate the position where the signal size was greatest. This is known as the {\em core} of the shower.
Modern analyses on fast computers still use essentially this method of core location.

It was quickly recognised that the cloud chamber was not a suitable apparatus for measuring the shower directions.  Although there was some subsequent use for anisotropy studies by Rothwell, Wade and Goodings \citep{Rothwell-56} in the UK and by Wayne Hazen and Fred Hendel at Chacaltaya
for checking shower directions, the small area, and relative difficulty of maintaining stable operation over long periods, made it unsuitable for many studies.

\section{Developments Arising due to the Availability of Photomultipliers}
\label{sec:PMT}
\subsection{Developments at the Culham Array and Consequences}

The increasing availability of PMTs led to some significant advances in the air shower technique. Two developments made at the Culham array led to the use of Cherenkov radiation to study extensive air showers.  The first -- which led directly to use of the Cherenkov technique to study high-energy photons (see \citep{Lorenz-epjh}) -- arose from Blackett \citep{Blackett-48b} pointing out the essential features of the production of Cherenkov radiation in air.  He calculated that the energy thresholds for electrons and muons at standard temperature and pressure were 20\,MeV and 4\,GeV, respectively.  He also showed that the Cherenkov radiation produced by cosmic rays traversing the atmosphere comprised $\sim 10^{-4}$ of the night-sky background so setting a limit to the darkness of the sky at night even under cloud.  He commented that ``presumably such a small intensity of light could not be detected by normal methods''.  This was Blackett's only publication on the topic but, Lovell relates \citep{Lovell-75} that he became interested in the quantum efficiency of the eye at this time, concluding that extensive air showers should produce a flash of light that he should be able to see lying down and looking upwards under suitable dark sky conditions, an investigation which Blackett carried out himself.  The outcome of Blackett's efforts seem not to have been recorded but his work inspired Galbraith and J V Jelley to search for flashes of light associated with extensive air showers using PMTs \citep{Galbraith-53}.

During moonless nights Galbraith and Jelley pointed a searchlight mirror of 25\,cm diameter and $\sim 12$\,cm focal length vertically with a PMT of 5\,cm diameter at the focus of the mirror.  The output from the PMT was connected directly to an oscilloscope and pulses were seen at a rate of about 1 per minute with the threshold set at three times that of the night sky noise.  Following this success the oscilloscope was triggered when Geiger counters of the Culham array were struck in coincidence and pulses were again observed.  While they established that the light pulses were associated with cosmic radiation, there was no evidence to support the hypothesis that the light observed was Cherenkov radiation. 

They continued their investigations the following year, moving to the Pic du Midi to take advantage of the greater number of nights of high clarity and in an elegant series of experiments they demonstrated that the light signals had the polarisation characteristic of Cherenkov radiation and had a spectral distribution consistent with what was predicted \citep{Galbraith-53}.  The detection of polarisation eliminated the possibility that the light was due to recombination radiation while calculations showed that the bremsstrahlung process gave insufficient photons to explain the observations.  In the same sequence of experiments they demonstrated a correlation of the light signals with shower energy estimating the threshold for detection as $\sim 10^{14}$\,eV.  

The incorporation of the detection of air-Cherenkov radiation came to be common at a number of shower arrays, but in the UK, which does not have an optimum climate for such work, the Culham activity was not followed up for nearly 20 years.  An important development was made by Chudakov and colleagues \citep{Chudakov-60} who measured the flux of Cherenkov radiation in air showers of $N \sim 10^5$-$10^6$.  These data were used independently by Greisen \citep{Greisen-56} and Nikolsky \citep{Nikolsky-62} to derive a relationship between the primary energy and the shower size by application of the track-length integral (see discussion in Sec.\,\ref{sec:Cornell}).

\begin{figure}[t]
\centerline{\resizebox{0.7\textwidth}{!}{\includegraphics{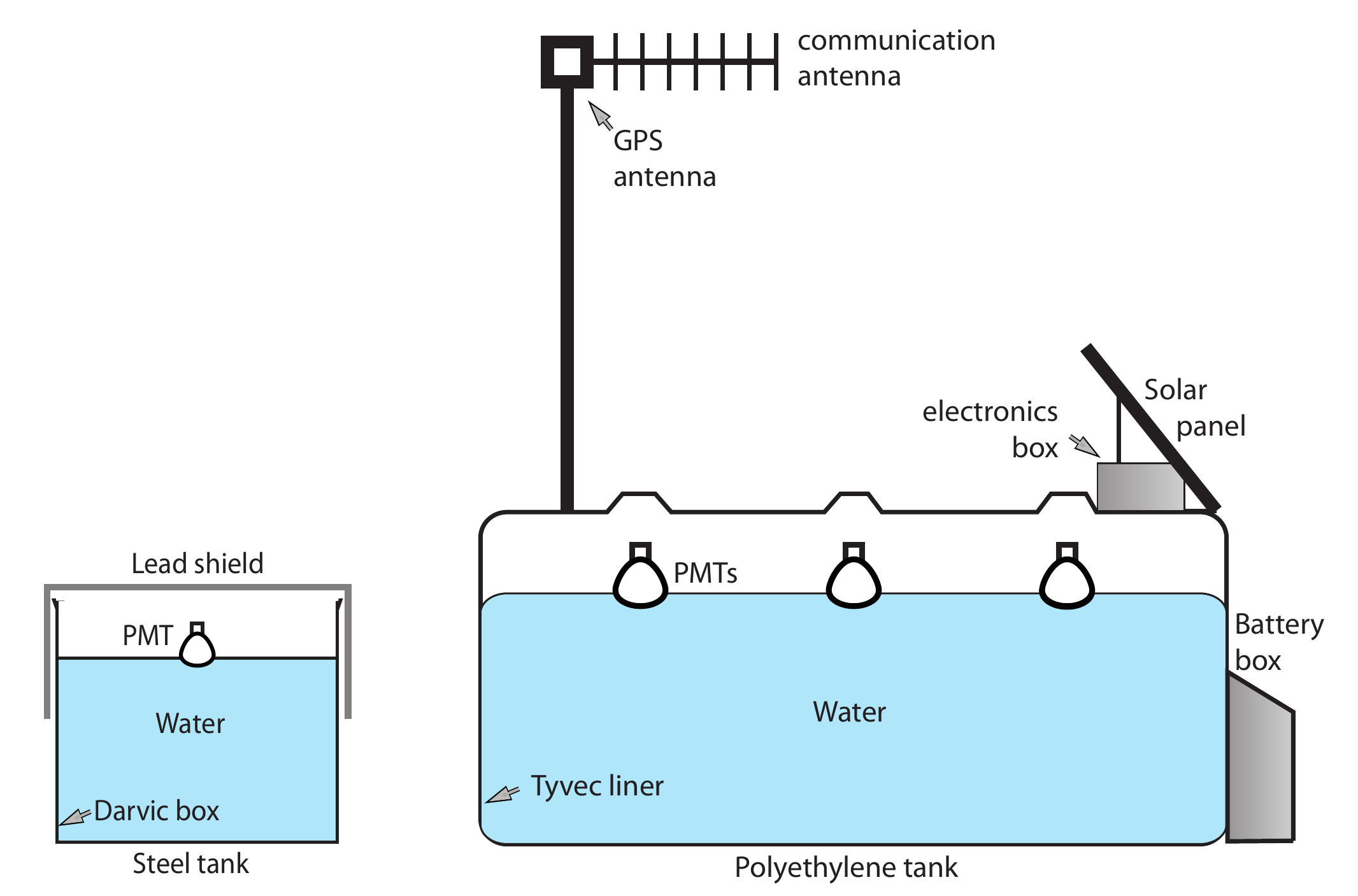}}}
\caption{
Scale comparison of the first water-Cherenkov detectors used by Porter et al.\ \citep{Porter-58} of 1.44\,m$^2$ read out by a single 5'' diameter PMT to those used at the Pierre Auger Observatory of 10\,m$^2$ read out by three 9'' PMTs \citep{auger-04}.}
\label{fig:Tanks}
\end{figure}

The second key development made at Culham was of the water-Cherenkov detector.  Credit for this work goes to N A Porter who, while a member of the team working with the Geiger-counter array, became the first to succeed in preventing bacterial growth in unfiltered water long enough to realise a stable detector \citep{Porter-58}.  This was achieved by enclosing the water in a cubical, sealed, box of `Darvic', a material then manufactured in the UK for use in sandwich boxes and therefore containing an inhibitor of bacterial growth.  Darvic was, however, chosen primarily for its white diffusive surface and its other properties only became known to the air shower community many years later.  The depth of water was 92~cm.  One of several advantages of a water-Cherenkov detector is that it enables the energy flow in the shower to be measured. 
Porter's detector can be seen as the prototype of those that were used at Haverah Park (1967-1987) and at the Pierre Auger Observatory (from 2000).  Indeed, there has been remarkably little advance over Porter's design, in which the PMT looked downwards into the water, to that of the present Pierre Auger Observatory \citep{auger-04} (see Fig.\,\ref{fig:Tanks}).

The Culham array was closed in 1958 to make space for the construction of the Culham Fusion Laboratories and in response to political moves towards 'useful research'.

\subsection{Developments by the MIT Group}

An extremely important development arising from the availability of PMTs was made at MIT under Rossi's leadership.  He had realised that the short fluorescence-decay times found in the newly-discovered liquid scintillators might make it feasible to construct large area detectors in which fast timing of the arrival of the particles of a shower would be possible.  The scintillating material chosen was a solution of terpenyl in benzene held in 5 gallon ($\sim 20$ l) drums of 600\,cm$^2$ cross-section.  Using three of these detectors, mounted in various configurations on the roof of the Physics Department at MIT, Bassi, Clark and Rossi \citep{Bassi:1953vl} showed that the particles in the disk of the shower were spread over a thickness of only a few metres. By shielding one of them with up to 20\,cm of lead, they demonstrated that the electrons in the shower lead the muons close to the shower axis.  The discovery that the shower disk was relatively thin ($\sim 10$\,ns) opened up the possibility of measuring the direction of the primary particle.  It is worth pointing out, in view of later discussion, that had the disk been thick (say $> 100$\,ns), sampling of the front with detectors as small as those used by Bassi et al.\ on a baseline of only a few metres would have greatly impaired the accuracy of reconstruction.  Assuming that the direction was perpendicular to a plane tangent to the surface defined by the leading particles in the shower, it was shown that the direction of the shower could be found to $\sim 2^\circ$.  This was a major advance over the crude collimating effect of the atmosphere. 

\begin{figure}[t]
\centerline{\resizebox{0.47\textwidth}{!}{\includegraphics{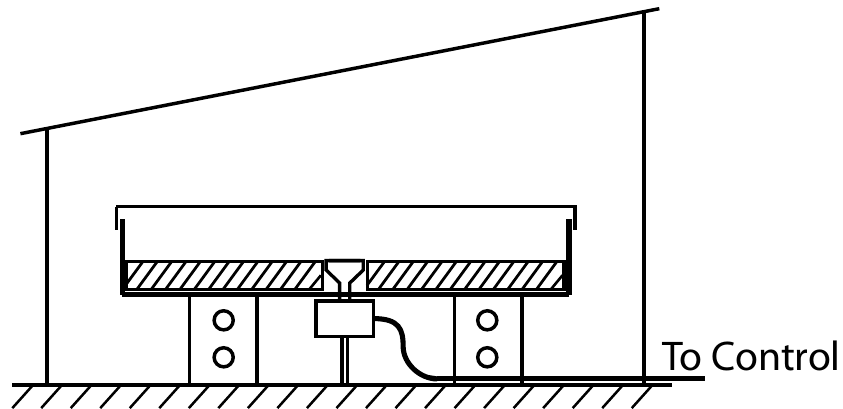}}}
\caption{
Schematic diagram of a scintillation counter used in the Agassiz shower array. The scintillator block was 105\,cm in diameter and 10\,cm thick.  The inside of the box was painted white and the diffuse light reflected from the walls was collected by a Dumont 5'' diameter PMT (Reproduction from \citep{Clark:1957vb}).}
\label{fig:Clark-Scintillator}
\end{figure}

This pioneering work led to the construction of a larger array at a partially wooded site, the Agassiz Astronomical Station of the University of Harvard.  Unfortunately the liquid scintillators were flammable and after a lightning-induced fire a method of making solid scintillator in large slabs with masses of $\sim 100$\,kg  was developed \citep{Clark:1957vb}.  These could also be viewed by PMTs and a schematic diagram of one scintillation counter is shown in Fig.\,\ref{fig:Clark-Scintillator}.

\begin{figure}[h]
\centerline{\resizebox{0.5\textwidth}{!}{\includegraphics{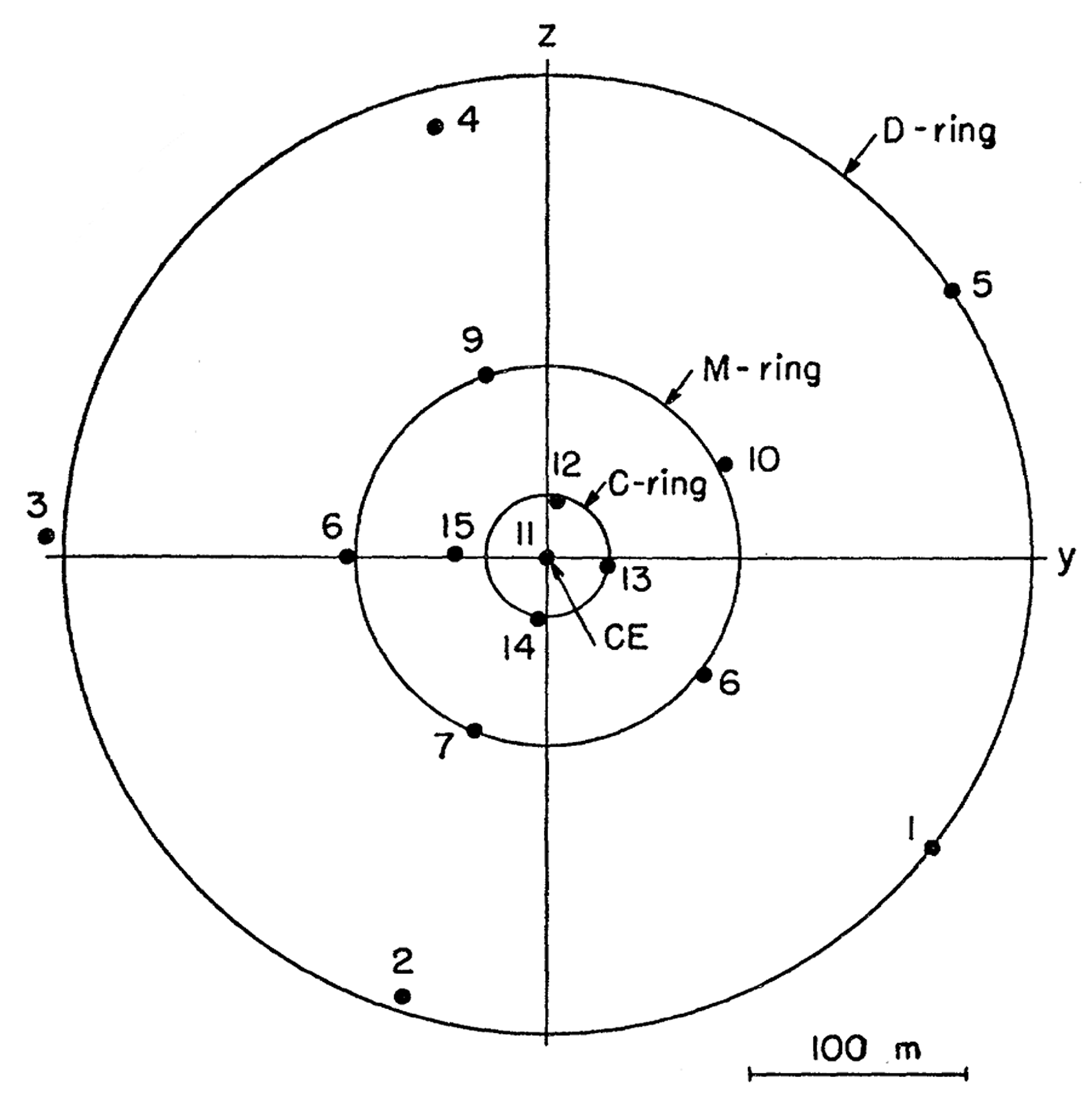}}}
\caption{
Schematic diagram of the MIT scintillation detector array. The four detectors in the C-ring were used only during a small part of the running time to extend the results to small shower sizes of $5 \cdot 10^5$ particles \citep{Clark:1957vb}.}
\label{fig:Clark-Layout}
\end{figure}

At the Agassiz site an array of 15 such detectors was operated between 1954 and 1957 with the layout shown in Fig.\,\ref{fig:Clark-Layout}.  Members of the group included George Clark, William Kraushaar, John Linsley, James Earl, Frank Scherb and Minoru Oda, who became a leading figure in air shower work in Japan.

An excellent first-hand account of Rossi's work at MIT has been given by Clark \citep{Clark-06}.

\subsection{Japanese-led Developments}

Cosmic-research began in Japan in the 1930s at RIKEN in Tokyo first under the guidance of Y Nishina and then under S Tomanaga.  At the end of WW\,II, experimental work in nuclear physics in Japan was essentially terminated for some years following the destruction of the cyclotrons at RIKEN and those in Kyoto and Osaka.  By contrast, cosmic ray work flourished: Tomanaga stimulated studies of extensive air showers at Mt Norikura (2770\,m)
\citep{Ito-97} and played a key role in establishing the Institute for Nuclear Studies (INS) in Tokyo.  He was also instrumental in encouraging J Nishimura and K Kamata to develop three-dimensional analytical calculations of electromagnetic cascades, work which they began after reading the Rossi and Greisen article of 1941 \citep{Rossi:1941ts} during daily visits to a US reading room in Tokyo.  Japan has been one of the leading countries in cosmic ray physics ever since.

The work at Mt Norikura, initially controlled by the University of Osaka, came under the direction of the INS.  Oda returned to Japan from USA in 1956 and played a major role in Japanese cosmic ray research before moving to $X$-ray astronomy.  Oda and K Suga became major figures on the Japanese cosmic ray scene.  Work at Osaka was led largely by S Miyake.

In Tokyo, at sea-level, Oda and his colleagues (notably T Matano, Suga, Y Tanaka and G Tanahashi) built up a series of shower arrays of increasing complexity leading finally to a configuration with 14 scintillators of 1\,m$^2$ area each, 5 smaller scintillators for fast timing, 4 scintillators of 2\,m$^2$ each installed in a tunnel 15\,m underground setting a threshold of 4.5\,GeV for muon detection. They were supplemented by 5040 neon tubes each of 2\,cm diameter spread over a small area of $2 \times 3.5$\,m$^2$ for muon counting -- a device that would now be termed a `calorimeter' -- to measure features of the hadronic component, and two total absorption Cherenkov detectors made of lead glass and a lead nitrate solution of $\sim 0.5$\,m$^2$ to study electromagnetic energy flow which was a feature of work at this time both in England and in Japan \citep{Fukui:1960tl}.

By far the most important insights came from the combined data from the 4 muon and 14 scintillator detectors.  Although the results have long since been surpassed, the INS group were the first to point out the key information that could be derived from a study of plots of muon versus electron number, $N_\mu$ vs.\ $N_e$ plots in modern language. 

\begin{figure}[h]
\centerline{\resizebox{0.65\textwidth}{!}{\includegraphics{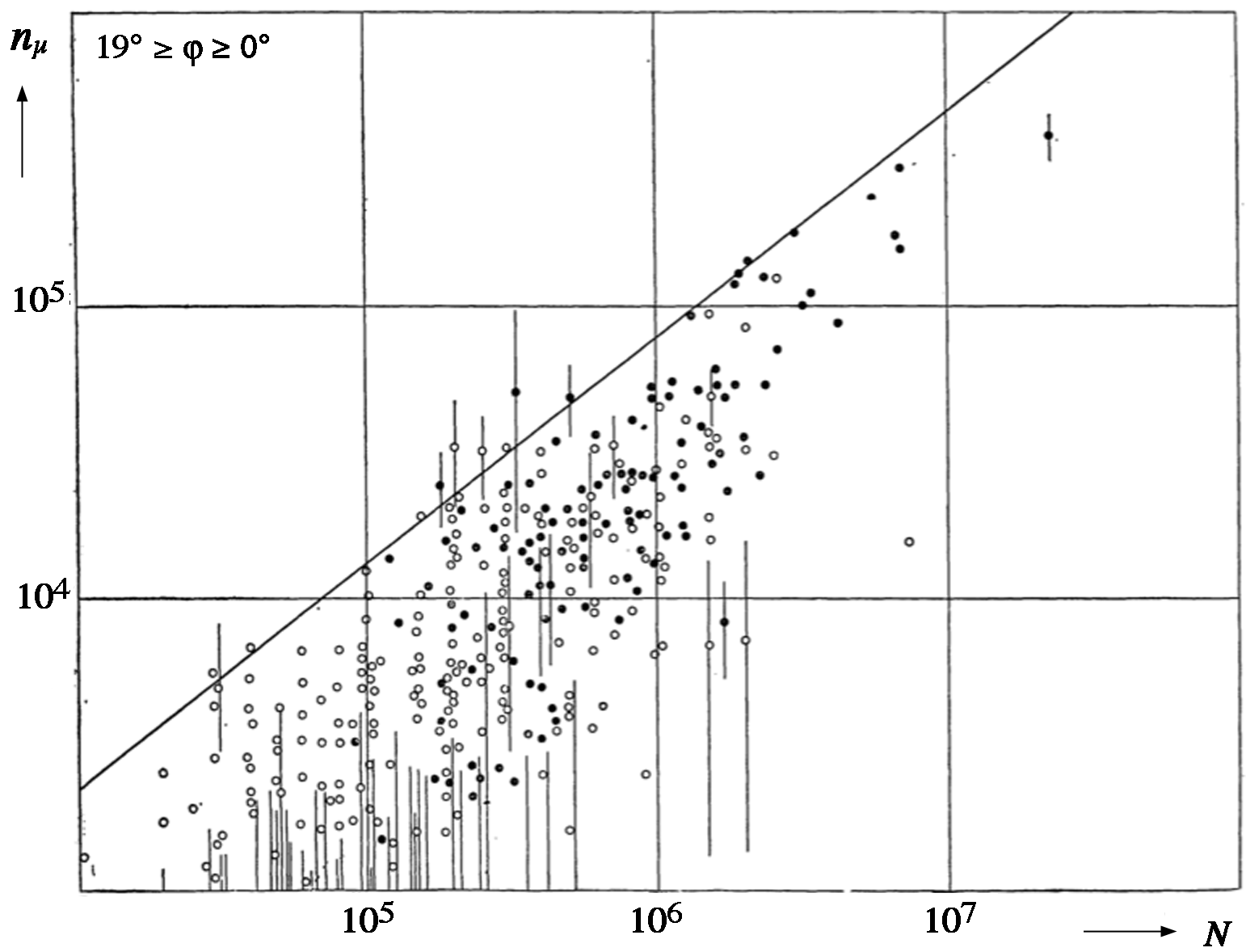}}}
\caption{
Reconstructed muon number $n_\mu$ vs.\ shower size $N$ for vertical showers as measured by \citep{Fukui:1960tl}.}
\label{fig:Fukui-nmu-ne}
\end{figure}

One of the plots from the INS work is shown in Fig.\,\ref{fig:Fukui-nmu-ne} for nearly vertical events.  Large fluctuations in muon number are evident for fixed electron size with the sharp upper boundary suggesting that the first interaction dominates the fluctuations in the shower.  This type of diagram, with improved statistics and smaller uncertainties in $N_\mu$ and $N_e$, when combined with detailed shower simulations, later proved to be a powerful tool for extracting information on primary mass.  In addition, it was soon recognised, when Monte Carlo studies developed, that at fixed primary energy the fluctuations in electron number were greater than those for the muons.  Accordingly the muon number came to be used as a proxy for shower energy.  The Japanese workers also gave important, though qualitative, analysis as to how details about heavy primaries could be extracted from such data and a number of studies were made subsequently which appeared to show anisotropies in the pattern of the arrival directions of events rich in muons.  The $N_\mu$ vs.\ $N_e$ plots also led to the idea of selecting showers with very low muon content as potential $\gamma$-ray candidates, a motivation behind the INS group joining the effort to found the air shower array built at Chacaltaya.


Two other innovations were made by the INS group: One was to attempt to observe the reflection of a radar signal from the ionisation column left by the shower in the atmosphere (Sec.\,\ref{sec:radar}). Additionally, Oda and Suga were amongst the originators of the fluorescence method of studying showers (Sec.\,\ref{sec:Fluorescence}).
The work at INS was also the forerunner of other projects in air shower research.  In addition to the BASJE project at Chacaltaya (Sec.\,\ref{sec:Bolivia}), the activities led to the Akeno and AGASA arrays in Japan and the Telescope Array in the USA. 



\section{The Impact of the MIT Group and of Rossi}
\label{sec:MIT}

The impact of the MIT group on the understanding of the extensive air showers was enormous.  As well as seminal technical advances, they introduced the analysis techniques that are the basis of methods that have been used to deal with data from surface arrays ever since.  The group was the first to develop routines to derive the direction of the shower from the fast-timing measurements and to find the position in the plane perpendicular to the shower where the signal size would be the largest, the shower core.  The MIT group was also the first to report a measurement of the differential shower size spectrum above $N = 10^6$ and took the first steps towards determining the energy of the primary particles using models of shower development.  

The determination of the energy that has created a particular shower is not straight forward and it is instructive to appreciate the various approaches that have been adopted over the years.  Although it was established relatively early that air showers contained nucleons, pions and muons in addition to an abundance of electrons and photons, the gross features of showers were found to be relatively well-described under the assumption that the primaries were electrons.  It thus became the practice to infer the primary energy from a measurement of the total number of charged particles, $N$, -- dominantly electrons and positrons -- in a shower, relating this to the primary energy using theory provided by such as the Nishimura-Kamata equations that describe the lateral distribution of charged particles for showers produced by photons or electrons.  The number of particles was straight forward to measure when the detectors were Geiger counters as they respond predominantly to charged particles.  Also, for the study of the showers produced by primaries of energy less than $\sim 10^{17}$\,eV, it was practical and economically feasible to build arrays in which the average separation of the detectors was less than the Moli\`{e}re radius\footnote{The 
Moli\`{e}re radius is the root mean square distance that an electron at the critical energy is scattered as it traverses one radiation length.}, about 75\,m at sea-level:  roughly 50\,\% of the charged particles of a shower lie within this distance.

As greater understanding of showers developed, there were moves away from using the photon/electron approximation to estimate the primary energy from the number of charged particles measured in the shower.  Also a difficulty in obtaining $N$ was recognised as scintillation counters were increasingly introduced during the 1950s.  Because of the success of the approach with Geiger counters and the lack of other methods to find the energy on an event-by-event basis, considerable effort was initially expended in relating the scintillator measurements to what would have been the particle count had a Geiger counter been located at the same position as the scintillation counter.  This adjustment to particle number was reasonable while the spacing between detectors remained small.  For example, at the Agassiz array, measurements were made at distances much closer to the shower core than one Moli\`{e}re radius (see Fig.\,\ref{fig:Clark-Layout}) and the scintillator response was converted to particle number using an array of Geiger counters operated for that purpose.  As more understanding of shower structure developed, the importance of the thickness of the scintillators was recognised and it was also realised that the conversion from scintillator signal to number of charged particles depended on the distance of the scintillators from the shower core as the energy spectrum of electrons and photons was distance-dependent.

To determine the energy spectrum from their measurements the MIT group made two innovations \citep{Clark-58}.  Firstly, they devised a method to relate the shower size observed at a zenith angle, $\theta$, to the size that it would have had had it come at $\theta = 0^\circ$.  The size, $N_v$, was defined as the vertical equivalent size such that 
\begin{equation}
N_v = N \cdot \exp [ X_0 (\sec \theta - \sec \theta_0)/\lambda ],
\end{equation}
where $X_0$ is the atmospheric depth in the vertical direction, $\theta_0$, is the angle to which the size is being referenced (usually $0^\circ$ in early work) and $\lambda$ is the attenuation length for showers in the atmosphere. This equation is derived from the observation that the size of showers above the same integral rate at different zenith angles decreases exponentially with atmospheric depth.  This approach to normalising shower size to a particular zenith angle is known as the {\em ``constant intensity cut method''}.

Secondly, they pointed out that to obtain an energy spectrum from the observed size spectrum required ``a quantitative knowledge of the cascade processes initiated by primary particles of different energies''.  This problem of quantitative knowledge of the hadronic process is still an issue over 50 years on though there is a growing understanding of the key hadronic interactions, most recently from the LHC.  S Olbert, one of the MIT group, had solved the shower equations to relate the shower size at different atmospheric depths to the primary energy \citep{Olbert-57}.  Using two models of high-energy interactions then current (the Landau model \citep{Belenki-56} and the Fermi model \citep{Fermi:1950uo,Fermi:1951wi}), making the assumptions of a collision mean free path for protons of 100\,\gcm ~and complete inelasticity, Olbert obtained relations between $N$ and energy, $E_0$.  For showers with $N = 10^8$ the Fermi and Landau models estimates were $3.13 \cdot 10^{17}$ and $3.55 \cdot 10^{17}$\,eV, respectively.

A study of the muon content of showers was made using a hodoscoped system of Geiger counters shielded with lead.  This work established that roughly 10\,\% of the particles in an air shower were muons \citep{Clark-58}.

A final report on the work of the MIT group was made in 1961 \citep{Clark:1961ws} where details of the largest event with a `Geiger-counter size' of $N=2.6 \cdot 10^9$ were given.  The array at Agassiz had been operated for about a year, from July 1956 to June 1957.  In addition the group had run a small shower array at Kodaikanal in India to search for anisotropies in the arrival directions at energies just above $10^{14}$\,eV.

The work directed by Rossi led to the establishment of the shower array at Chacaltaya (with the Japanese group from INS as major partners).  A particular motivation was to search for $\gamma$-rays by attempting to identify showers containing fewer muons than average.  This attempt was unsuccessful but the first indirect deductions about the position at which the number of particles in showers of $\sim 10^{16}$\,eV reach their maximum were obtained.
Additionally, attempts were made to find the depth of shower maximum using the constant intensity cut method, a very important technical conception.  Rossi also encouraged the work led by Linsley at Volcano Ranch in New Mexico to establish the first array with an area of over 1\,km$^2$, built to study the highest energy events (Sec.\,\ref{sec:Volcano}).

\section{Work of the Cornell Group and the Influence of Greisen}
\label{sec:Cornell}

Greisen, a former student of Rossi, who had also worked at Los Alamos, founded a group at the Cornell University during the 1950s.  The first step was to build an array of radius 500\,m with $15 \times 0.85$\,m$^2$ scintillators with 5 near the centre on 3 to 80\,m spacing and 5 each 150 and 500\,m from the array centre.  Like the MIT group, the Cornell team did not have a fast computer available to them initially and developed some ingenious analogue methods to find the direction and the shower core. This could only be used for a relatively small number of large events.

The scintillation detectors were surrounded by 10 to 15\,cm of insulation as a major aim of the project was to study the anisotropy of cosmic rays down to energies as low as $10^{13}$\,eV where temperature effects in the apparatus could mimic sidereal anisotropies.  Above $10^{13}$\,eV, around $10^4$ events were recorded per day with several million events accumulated in 1957 and 1958.  The results of this study remain important because of the magnitude of the sample and the care with which atmospheric effects were treated.  A striking result was the observation of a remarkable similarity of the phase \citep{Delvaille-62} close to 20 hours in right ascension, seen in many of the contemporary experiments.  In the absence of direction measurements, what could be said was that the showers had a distribution in declination that is approximately Gaussian with a width of $\sim 20^\circ$ and peaked at $40^\circ$ (the latitude of Cornell is $42.5^\circ$\,N).

An important finding was that the average radius of curvature was 3300\,m, with it exceeding 10\,km in some events.  This implies that a correction for the curvature was necessary when determining the direction of showers in which the core was far from the centre of an array \citep{Benett-62}.

A measurement of the size spectrum above $N = 6 \cdot 10^6$ was made using an approach similar to, but independent of, that of the MIT group.  Expressing the results as $K(N/10^6)^{-\gamma}$ the values of $K$ reported for Cornell and MIT respectively are $3.58 \pm 0.28$ and $3.48 \pm 0.53$, and for $\gamma$, $1.90 \pm 0.10$ and $1.84 \pm 0.6$.  Thus, the two measurements were found to be in good agreement.  The largest event recorded with the Cornell array contained $N \simeq 4 \cdot 10^9$ particles.

Greisen and his group also studied muons in showers, extending what had been done at MIT and elsewhere.  They were the first to use a magnetic spectrometer to determine the momentum of muons and indeed only two other magnetic spectrometers were constructed subsequently for use in air showers. These were built by the groups at Haverah Park and at Moscow State University \citep{Earnshaw:1967vr,Vashkevich-80}.
%
The spectrometers were precision instruments but the relative insensitivity of muon momentum spectra to features of air shower development and the relative complexity of operation caused interest in their use to fade although each of the devices were used to deduce important results.

From their measurements, the Cornell group derived useful formulae to describe the lateral distribution of muons above 1\,GeV and also the energy spectrum of muons as a function of distance.  Although the muon sample was only 559, and the shower analysis was not done on an event-by-event basis, the relations established have been found to fit a wide sample of modern work on the muon lateral distribution even for showers of greater energy. 

In his seminal reviews \citep{Greisen-56,Greisen-60} Greisen worked out parameterisations of both the electron and muon lateral distributions (LDF) which described the data well over a large range of atmospheric depths and distances from the shower core. 
He also noted that his parameterisation of the electromagnetic distribution was a close approximation to the analytical calculations for electromagnetic showers performed by Kamata and Nishimura \citep{Kamata-58}. Greisen's approximations to the Nishimura-Kamata functions for shower ages $0.5 < s < 1.5$ has become known as the Nishimura-Kamata-Greisen (NKG) function.
He also suggested a functional form for the muon LDF
now referred to as the Greisen-function.
Both parameterisations became widely used in analyses of surface detector arrays even though modifications optimised to specific experimental conditions were devised later.

The Cornell group also found that the mean energy of muons in showers of $N \sim 10^6$ particles ($E_0 \sim 10^{16}$\,eV) was 7\,GeV.  Furthermore, they showed that the energy spectrum of muons in showers was significantly flatter than that of unassociated muons and that the excess of positive muons $\eta = (\mu^+ - \mu^-)/(\mu^+ + \mu^-) = - 0.023 \pm 0.044$.  The latter result is of importance as $\eta = 0.11$ for unassociated muons.  The positive excess for unassociated muons can be explained as demonstrating an excess of protons over neutrons in the primary flux or from a charge asymmetry in kaon and hyperon production.  That $\eta \simeq 0$ for showers allows these two possibilities to be separated and the Cornell team were able to conclude that in the energy range studied, kaon production is an order of magnitude less frequent than pion production.
The measurements of $\eta \simeq 0$ were confirmed with better precision by the Haverah Park and Moscow State University groups  \citep{Earnshaw:1967vr,Vashkevich-80}.
An important by-product of this spectrometer work at Haverah Park was the comparison made between the directions of muons with no detectable deflection and the shower direction deduced from fast-timing of the arrival of the signals recorded in the water-Cherenkov detectors. 

The solid-iron spectrometer of the Moscow State University array was operated under 40\,m.w.e., yielding a threshold for muons above the ground of 10\,GeV.  It had the highest maximum detectable momentum of any of the three instruments used.
The spectrum and lateral distribution were in agreement with the quark-gluon-string (QGS) model of the time \citep{Vashkevich-80}.  An important deduction was the fraction of energy of the shower that was carried by the muon component: this was found to be $(13 \pm 3)$\,\% for muons above 1\,GeV for primaries of $10^{16}$ - $10^{17}$\,eV.

After work on the Cornell scintillator array had been completed, Greisen turned his attention to the development of the fluorescence technique.  His two reviews \citep{Greisen-56,Greisen-60} remain important sources of insights. In particular, in the first he developed a method to estimate the energy that a primary particle would need, on average, to produce a shower of a certain size.  

It had long been realised that most of the energy of the primary particle was dissipated in ionisation
of the air. Apart from a small fraction $F(E_0)$ of energy lost to neutrinos and high energy muons entering into ground, the primary energy $E_0$ is given by the track-length integral,
\begin{equation}
(1-F)\cdot E_0 \simeq \alpha \int_0^\infty N(t) dt,
\end{equation}
where $N$ is the number of particles at a depth $t$ and $\alpha$ is the energy loss per unit of path length.  If $t$ is expressed in radiation lengths then the energy of the primary particle is given by multiplying the value of the integral by $\epsilon_c$, where $\epsilon_c$ is the critical energy, (84.2\,MeV for air).  It is assumed that $\epsilon_c$ is an accurate measure of the average energy lost per radiation length by the charged particles of the shower.  A general idea of the form of $N(t)$ can be gained by inspection of Fig.\,\ref{fig:cloud-chamber} but, while in the case of the cloud chamber event the integral can be evaluated relatively directly, Greisen's first evaluation was made using measurements of the rate of detection of showers made at sea-level, at mountain altitudes and in airplanes to obtain the shape of the longitudinal development curve.  Combining measurements made with different instrumentation by many people was not easy and was confined to showers which had $\sim 10^5$ particles at sea-level.   He deduced that the number of particles in the shower at maximum would be $\sim 8.5 \cdot 10^5$ and be at a depth of 450\,\gcm.  The value derived for the track-length integral was 11.5\,GeV per electron at sea-level.

It is necessary to account also for energy deposited in other forms than in ionisation and Greisen made estimates of these, including muons, neutrinos, low-energy nucleons and nuclear excitations which increased the value to $14 \pm 3$\,GeV per electron.  He deduced the integral rate for primaries of $E > (1.4 \pm 0.3) \cdot 10^{15}$\,eV as $(1.7 \pm 0.3) \cdot 10^{-6}$\,m$^{-2}$\,sr$^{-1}$\,s$^{-1}$ which agrees with a modern estimate of the flux of $7.8 \cdot 10^{-7}$\,m$^{-2}$\,sr$^{-1}$\,s$^{-1}$ within about a factor of 2.  The depth of shower maximum at this energy is still not well-known but it is probably within 10\,\% of the 450\,\gcm\, from Greisen's study.

Greisen made a further estimate of the relationship between the observed number particles and the primary energy \citep{Greisen-60} building particularly on the studies of Cherenkov light made by Chudakov and his collaborators \citep{Chudakov-60} in the Pamirs.  He was also able to make use of new measurements of the energy and number of muons in showers and of the electromagnetic energy flow.  Chudakov and his group had shown that in a shower of $1.4 \cdot 10^6$ particles measured in the Pamirs, there were $1.2 \cdot 10^5$ photons from Cherenkov radiation.  Assuming that the bulk of this radiation was from electrons of $> 50$\,MeV, Greisen found that the total ionisation loss, above the observation level, for a shower of this size was $5.2 \cdot 10^{15}$\,eV or 3.7\,GeV per particle.  To get the total energy, 0.2\,GeV per electron has to be added for the each of the electromagnetic and nuclear particle components with a further 0.4\,GeV per particle for the muons and neutrinos giving the primary energy as $6.3 \cdot 10^{15}$\,eV.

While it is clearly not feasible to make such precise energy analyses for the components of much larger showers, as measurements close to the shower axis are not practical, the track-length integral method is the essence of the fluorescence technique (see Sec.\,\ref{sec:Fluorescence}) which Greisen did much to promote and, through measurements of the Cherenkov radiation, remains a key technique in the energy determinations at the Yakutsk, Tunka, and other arrays.

\section{Work in Bolivia}
\label{sec:Bolivia}

Of the several laboratories to be developed for the study of air showers, one of the most important, and certainly the highest, was constructed at Mt Chacaltaya in Bolivia at 5200\,m (530\,g\,cm$^{-2}$) and is still in operation.  The mountain had already been used extensively for the exposure of nuclear emulsion plates in the 1940s.  At Chacaltaya important steps were taken to infer the depth of shower maximum, to measure the energy spectrum and to study the mass of cosmic rays, including a search for photons.

As a first step to understanding the features of showers at high altitude, 11 of the scintillators used in the Agassiz experiment were deployed in an array of 700\,m diameter on the Altoplano at El Alto, near La Paz, Bolivia, at an altitude of 4200\,m (630\,g\,cm$^{-2}$) in 1958.  Showers of size $\sim 10^7$ were studied.  It was found that, unlike those of a similar size at sea-level, the steepness of the lateral distribution changed with zenith angle, being steeper for the more vertical showers.  Furthermore, for $N \sim 3 \cdot 10^6$ the change in shower size with depth from 630 to $\sim 800$\,g\,cm$^{-2}$ was small suggesting that these showers had their maxima close to 630\,\gcm~ \citep{Hersil:1961uv,Hersil:1962ve}.

In 1958, following a proposal by Satio Hayakawa, the MIT, INS and La Paz groups joined forces to establish the Bolivian Air Shower Joint Experiment (BASJE) at Mt Chacaltaya which started taking data in the early 1960s.  The program originally had two foci: one was to extend understanding of the longitudinal development using the ideas of the constant intensity cut method and so measure the primary energy spectrum while the other was to make a search for photons of very high energy.  The basic shower array comprised the 20 Agassiz-like scintillators deployed within a circle of 150\,m diameter with 5 scintillators for fast timing, supplemented with a muon detector of 60\,m$^2$ array.  The muon detector was constructed from 160 tonnes of galena (the natural mineral form of lead sulphide) which was readily available locally.  Modules of 4\,m$^2$ commercial scintillator were developed by Suga and were used together with a logarithmic time-to-height amplifier \citep{Suga-61} to measure the muon flux in showers.  The 60\,m$^2$ of scintillator were placed below a concrete structure supporting the galena.  The size of this muon detector exceed those built previously by about an order of magnitude and made practical a search for showers produced by primary $\gamma$-rays under the hypothesis that such shower would have low numbers of muons.  Events with less than $10$\,\% of the average number of muons were found but they were not clearly separated from the bulk of the data and did not show any anisotropy. 

In addition to the energy spectrum measurements and the photon search, two innovative studies of the mass composition of cosmic rays, which seem to be over-looked in recent reviews of the field, were made.  S Rappaport and H Bradt \citep{Rappaport:1969ua} used the muon detector to measure the component of nuclear-active particles in showers and deduced that the composition at $\sim  10^{15}$\,eV was similar to that measured directly at $10^{11}$\,eV by making a comparison of the data with their three-dimensional Monte Carlo calculations \citep{Bradt:1967ut}.  Further, A Krieger and Bradt \citep{Krieger:1969ta} augmented the scintillator array with 9 open PMTs to detect air-Cherenkov light and concluded that at $\sim 10^{16}$\,eV the composition was much as it was at $10^{12}$\,eV.

\section{Early Ideas about Fluctuations}
\label{sec:fluctuations}

It was recognised early in the study of air showers that the calculation of average values alone was insufficient to interpret observations.  When it was thought that the primary cosmic rays were photons some progress on the problem was made by analytical methods but real advances awaited the detailed Monte Carlo calculations possible with the advent of adequate computing resources.  Early work concentrated on electromagnetic cascades and, while at shower maximum the fluctuations due to the number of particles is negligible when $N \sim 10^6$, this is not the case either before or after shower maximum when the fluctuations can be very large.  For example Greisen \citep{Greisen-56} showed that for a shower of this size near sea-level the number of particles might be increased in an electromagnetic cascade by $\sim 50$\,\%.  L Janossy and H Messel \citep{Janossy-50} established that at shower maximum the mean deviation was $N_{\rm ave}^{1/2}$ where $N_{\rm ave}$ is the average number of particles at shower maximum, while, according to Greisen (1956), Bhabha and Ramakrishnan found this to be $\sim N_{\rm ave}$.  In one of the first applications of the Monte Carlo technique the results of Wilson \citep{Wilson:1952wc} were found to support the conclusions of Janossy and Messel.

The radiation length for an electron at $\sim 37$\,\gcm\, is smaller than the mean free path for interaction of protons ($\sim 80$\,\gcm\, at low energy) or pions ($\sim 120$\,\gcm).  Thus for protons the average number of interactions in a vertical traversal of the atmosphere is $\sim 12$-15.  It follows that fluctuations in the position of the first interaction in the atmosphere and the way in which the shower grows can be large.  It also follows that fluctuations in showers created by heavier nuclei will be smaller than for proton showers though the exact detail of the break-up of a nucleus in the first collision is clearly important.  The simple approximation is that an iron shower is caused by the superposition of showers produced by 56 nucleons though more sophisticated descriptions of the process were made as more data, particularly from nuclear emulsion studies, became available.

Before the advent of Monte Carlo calculations, the analytical calculation of fluctuations was very difficult.  Nonetheless some insights were made and as sometimes happens such insights occurred to different people at about the same time.  The analyses of Zatsepin \citep{Zatsepin-60} and of Cranshaw and A M Hillas \citep{Cranshaw-60} are instructive and came to similar conclusions.  By the late 1950s it was recognised from many observations that some properties of showers did not change with energy or altitude or zenith angle in the manner that first considerations led one to believe.  For example the attenuation length of shower particles can be deduced from the way in which the rate of showers recorded varies with barometric pressure.  The barometric coefficient, already noted by \citep{Auger:1939wp} and first reported in \citep{Auger:1942tm}, is given as $1/\Lambda = - \partial \ln R/\partial t$, where $R$ is the shower rate and $t$ is the atmospheric depth.  It can be shown that if $\gamma$ is the slope of the integral spectrum of shower sizes that $\lambda = \gamma \Lambda$, where $\lambda$ is the attenuation length of the shower particles.  It was found that $\Lambda$ decreases as the shower size increases and this, combined with the increase of $\gamma$ with shower size, leads to $\lambda$ being approximately constant.  From a number of experiments for showers of size $10^4$ to $10^6$ it was found that $\lambda$ was $\sim 180$\,\gcm.

\begin{figure}[h]
\centerline{\resizebox{0.35\textwidth}{!}{\includegraphics{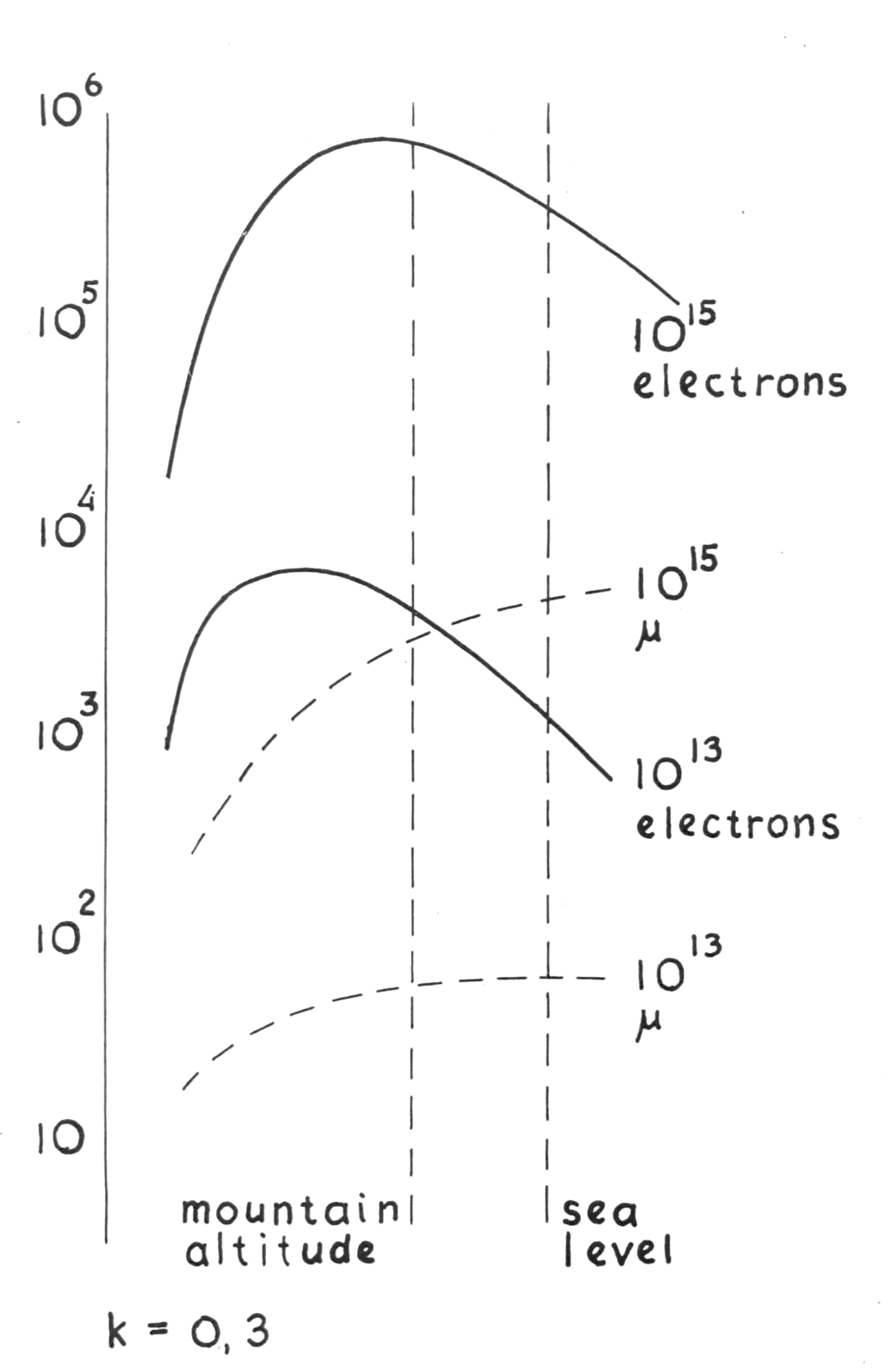}}}
\caption{
Altitude dependence of the average electron and muon particle numbers in showers for primary energies of $10^{13}$ and $10^{15}$\,eV assuming an inelasticity $k=0.3$ \citep{Cranshaw-60}.} 
\label{fig:Cranshaw-Hillas-60}
\end{figure}

This result appeared to be consistent with the fact that the shape of the lateral distribution averaged over many showers was constant for all sizes and altitudes.  Only when it became possible to observe showers close to their depth of maximum, as at Chacaltaya, did this picture change.  However, such constancy is hard to understand if only average development curves are considered.  In the Fig.\,\ref{fig:Cranshaw-Hillas-60} 
it is clear that the attenuation length, the slope of the shower development curve at the observation depths would be expected to change with size or with altitude.

Also, it is evident from consideration of these average curves that the ratio of muons to electrons would be expected to change rapidly with altitude or zenith angle, contrary to what had been found.  Zatsepin and Cranshaw \& Hillas came to the same conclusion, namely that the shower cascade from each nucleon coming from a collision is rather short so that at any observation level only the particles of the last cascade, occurring relatively close to the observation level, are observed.  The length of the cascade is determined by the $\pi^0$s created in the collisions.  Zatsepin described the shower as being like a fir-tree whose branches fluctuate in number and strength.  An essential input to this discussion, in addition to the observations on the barometric coefficient and the shape of the lateral distribution, is realisation that collisions of the proton with a nucleus are elastic so that the nucleon retains a large fraction of the energy.

The idea that one cascade dominates at the observation level is less tenable when the energy increases beyond that needed to produce showers with $N \sim 10^6$.  At energies above $10^{17}$\,eV several generations of cascades contribute to what is observed.

Further understanding of fluctuations in showers came as the Monte Carlo technique was developed. However, because of the direct and intuitive relation between shower-to-shower fluctuations and the primary mass, the investigation of the EAS muon number fluctuations was historically the first method employed to study the primary cosmic ray mass composition \citep{Fukui:1960tl,Khristiansen-63}.

\section{Nuclear Interactions and Mass Composition from Studies of Shower Cores}
\label{sec:Cores}

It would be surprising if the lines of attack to the solution of the coupled problems of understanding particle physics at high energies and simultaneously obtaining the mass composition had all been successful and of course this was not the case.  It is evident, with hindsight, that the approach to tackling the mass problem via the $N_\mu$ vs $N_e$ route, suggested by the INS group in 1960, failed because none of the installations built before the 1990s was sufficiently large, particularly in terms of the area of muon detectors and thus the accuracy with which the muon number could be determined.  However, it is also true that Nature was unkind.  Up until 1973 it had been believed that the cross-section for $pp$ collisions had reached an asymptotic value but when it was found, from the ISR\footnote{The Intersecting Storage Rings (ISR) was the world's first hadron collider, operated from 1971-1984 at CERN.} measurements at CERN, that the cross-section increased with energy the problem of separating proton and iron primaries became harder.  
Parameters such as the lateral distribution function, the rise time of the signal and the radius of curvature of the shower front are essentially determined by the geometry associated with the shower development and, with the cross-sections getting larger as the energy grows, discrimination becomes harder because of the importance of the height of the first and other early interactions.

Before accelerator data were available, there was much speculation as to the nature of the nucleon-nucleon interaction.  Some of the best theorists of the post-war period (Fermi, Heisenberg, Landau and Oppenheimer) spent time discussing this question.  In particular it was unclear as to whether all of the energy in the centre-of-mass of the collision was radiated into secondaries from a hot plasma or if, as Lewis, Oppenheimer, and Wouthuysen proposed \citep{Lewis-48}, a process analogous to electron bremsstrahlung took place.  Under this latter picture multiple emission of pions was predicted to occur at narrow angles with respect to the axis of the collision and studies of the central region of showers, within a few metres of the shower axis, were undertaken to see if related effects could be found.  

Studies in the 1950s of the central region of the shower were made using relatively small area detectors ($\sim 0.5$\,m$^2$) of high spatial resolution: the results were inconclusive.  By the mid-1960s devices to study the cores of showers at energies above $\sim 10^{14}$\,eV had been developed at INS (spark chambers, 20\,m$^2$), Mt Norikura, Japan (12\,m$^2$, scintillators under 2\,m of water), Sydney (10\,m$^2$, plastic scintillator), Kiel, West Germany,(32\,m$^2$, neon hodoscope) and Tien Shan, USSR (121\,m$^2$, sparse array of scintillators).  As detectors of larger area were developed so was a better understanding of the role of fluctuations in the development of showers and it became apparent that the structure of the core region depended on the energy and mass of the incoming nucleus and on the nuclear cascade process and its fluctuations.  It is now clear that it was rather ambitious to hope to separate these effects.  Further, strong differences between what was observed in the different detectors were soon found.  An effect noticed early-on was that the presence of relatively light, but non-uniformly spread, material above the detectors, such as beams supporting roofs, could give rise to spurious claims of structure.  The relatively primitive power of Monte Carlo calculations was a further handicap at this time.

Unfortunately, some rather extreme claims were made for the potential of the detectors to produce data relating to the mass of the primary and of the transverse momentum that was associated with some of the secondary peaks observed.  For example, the Sydney group claimed that on their $64 \times 0.16$\,m$^2$ array of scintillators, showers with two cores could be identified due to primary deuterons and a flux was quoted above $10^{15}$\,eV \citep{Bray-66}.  In the discussion following this presentation, Hillas \citep{Hillas-66b} pointed out that fluctuations in the interactions of the 4 nucleons of a helium nucleus, freed in the first collision, would be large and that on average the third most energetic nucleon would retain only 6\,\% of the total energy at sea-level: many helium nuclei would thus be expected to give showers with two cores.  Claims such as this, plus reports from many groups of frequent examples of events showing evidence for transverse momenta much in excess of what was seen at accelerator experiments earned the field of cosmic rays a poor reputation. 

One of the cleanest experiments was that done by the group at the University of Kiel, under the direction of Tr\"umper.  As a result of their work it became possible to understand many of the differences between different arrays and to recognise that the problem of extracting mass composition and particle physics from the core-study approach was probably impractical. Insights in the later stages of this work came as a better understanding of fluctuations came from the use of Monte Carlo calculations.

Data taking at Kiel continued after Tr\"umper's move to T\"ubingen in 1969 and led to an important but puzzling claim which stimulated much activity (Sec.\,\ref{sec:Cygnus}).

\section{The First Surface Detector Arrays covering more than 1 km$^2$}
\label{sec:km2arrays}

Many small arrays were built to study the cosmic rays in the region from $10^{14}$ - $10^{17}$\,eV at locations across the world with scientists in Australia, Germany, India, Italy, Japan, Poland, UK, the USA and the USSR making important contributions.  The early measurements have been replicated with very superior statistics in the modern arrays built in Germany (KASCADE and KASCADE-Grande), in Italy (EAS TOP) and in Tibet: this applies particularly to the energy region $10^{14}$ to $10^{16}$\,eV which includes the region where the energy spectrum steepens. We shall discuss those briefly in Sec.\,\ref{sec:Current}.

By contrast the number of devices constructed with collecting areas of over 1\,km$^2$ has been only 7, including the Auger Observatory, the Telescope Array and the Yakutsk array that are still operating, although with the latter reconfigured to study smaller showers. 
A Soviet proposal for a 1000\,km$^2$ array named EAS-1000, led by Khristiansen, was given formal approval and construction began \citep{Anonymous:1989wz}, but the project was hit by the political and economic problems that came with the glasnost and perestroika and was never realized.
Data from the Telescope Array and the Auger Observatory currently dominate from the Northern and Southern Hemispheres, respectively.  By contrast to the low-energy arrays, it is useful to discuss the pioneering large arrays in some detail first, as at each different features of technique and analysis were introduced which were important for later studies.  The layout of these arrays can be found for example in the review by M Nagano and A A Watson \citep{nagano00b}: essentially all arrays are variations of the style developed at MIT shown in Fig.\,\ref{fig:Clark-Layout}.  While methods of data recording evolved, the analysis techniques were similar to those introduced at MIT.  

\subsection{Volcano Ranch}
\label{sec:Volcano}

The first of the giant shower arrays was constructed at Volcano Ranch, New Mexico (1770\,m, 834\,\gcm) by members of the MIT group \citep{linsley-61}. It consisted of 19 plastic scintillation counters of 3.3\,m$^2$ area, each viewed with a 5'' PMT.  For some of the operating time, the detector spacing was 442\,m which was increased to 884\,m, enclosing an area of 8.1\,km$^2$ for the major period of operation of $\sim  650$ days. In addition to the 19 units that were used to detect charged particles, a further unit of 3.3\,m$^2$ was shielded by 10\,cm of lead to give the muon density above 220\,MeV in some events.  The construction, maintenance and data analysis of Volcano Ranch was the almost single-handed effort of Linsley who made many contributions to the understanding of giant showers.

Data from this array yielded the first measurement of the energy spectrum of cosmic rays above $10^{18}$\,eV, giving the earliest hint of a flattening of the spectrum in that region \citep{Linsley-63a}, a hint that took over 20 years to confirm convincingly.  Linsley also made the first exploration of the arrival direction distribution of these exceptional events.  The most energetic one was assigned an energy of $10^{20}$\,eV \citep{Linsley:1963uo}, an energy that was subsequently revised to $1.4 \cdot 10^{20}$\,eV \citep{Wada-86}. This event, reported before the discovery of the 2.7\,K cosmic microwave background radiation and the subsequent prediction of a steepening of the spectrum, remains one of the most energetic cosmic rays ever recorded. 

The large spacing of the detectors at Volcano Ranch presented difficulties for determining the number of particles in the shower.  At 834\,\gcm\, the Moli\`ere radius is $\sim 100$\,m and for two very large events of energies $10^{19}$\,eV \citep{linsley-61} and $10^{20}$\,eV \citep{Linsley:1963uo} no measurements were obtained within 300\,m of the shower axis. It was argued that the size, $N$, could be inferred by integrating an average lateral distribution over all distances with the energy being deduced under the assumption that the showers were at the maximum of their development at the depth of observation.  The conversion to primary energy was based on Greisen's method of the track-length integral.  It is now known from direct measurements that the average depth of maximum, $X_{\rm max}$, is $\sim 100$\,\gcm\, higher in the atmosphere than then believed so that some of the inferred energies made using this method by the MIT and Volcano Ranch groups were possibly too low.

An important phenomenological investigation of the arrival time distribution of charged particles and of muons in showers \citep{Linsley:1962uw} was made establishing that at distances beyond 350\,m, the mean arrival time of muons was substantially earlier than that of all charged particles: at 350\,m the means are 40 and 160\,ns, respectively while at 1100\,m the corresponding figures are 380 and 700\,ns.  These data are useful for understanding the precision with which the shower front can be measured and the direction of the primary determined, and have also led to work with detectors of larger area targeted at finding the longitudinal development of showers.

To explore the lateral distribution of large showers in more detail, Linsley \citep{Linsley-73} split the detectors into 80 units of 0.815\,m$^2$ and arranged them with a spacing of 147\,m over an area of $\sim 1$\,km$^2$ in diameter.  He found that previous measurements of the change of steepness with energy were in error with too rapid a variation having been claimed.  This result led him to point out \citep{Linsley-77-elong} that, for a fixed mass composition, the shower maximum could not penetrate into the atmosphere more rapidly than $2.3 X_0$ per decade of energy, where $X_0$ is the radiation length.  While this result is anticipated in the Heitler model of an electromagnetic cascade, Linsley turned this into an important tool for organising data and for testing whether calculations of shower development that were becoming more numerous at this time were correct. The rate of change of the depth of maximum with energy is known as the {\em ``elongation rate''}, a term introduced by Linsley and widely adopted.  It is closely related to the multiplicity of production of secondary particles in the shower: the limiting value corresponds to the case where a single $\pi^0$ is produced in the first interaction and takes all of the energy of the primary particle so that the cascade develops at a rate governed by that associated with a single photon, essentially as given by Heitler's simple model (Sec.\,\ref{sec:BasicIdeas}).

\subsection{Haverah Park}
\label{sec:HP}

Following the closure of the Culham array in 1958 it was decided, under the strong influence of Blackett, that work on extensive air showers should continue in the UK but be supported and developed within the university environment by a team drawn from several universities. This led to the construction of the Haverah Park array (1964 - 1987) under the leadership of J G Wilson until his retirement in 1976, with strong support in the initial stages from R M Tennant.  Prototype studies were carried out at Silwood Park near London under H R Allan who led a small team to examine the potential of the Cherenkov detectors developed by Porter at Culham \citep{Allan:1962vk} and A W Wolfendale who led an effort to evaluate the potential of neon flash tubes.

While the Silwood studies were underway, a site search identified land about 25\,km from the University of Leeds  (200\,m)
where an array covering 12\,km$^2$ was established and which operated for 20 years from 1967 to study features of showers from $10^{15}$ to $10^{20}$\,eV.  Restrictions on land access made it impossible to position the detectors on a uniform grid.  The solution adopted was to have a central, four-detector water-Cherenkov (34\,m$^2$) array with 500\,m spacing, together with six sub-arrays of 50 and 150\,m spacing $\sim 2$\,km from the centre.  Arrays of 50 and 150 m were also centred on the central detector to enable studies of low energy events.  The primary detectors were water-Cherenkov detectors of $2.25\,{\rm m}^2 \times 1.2$\,m with over 200 being deployed.   In addition there was 10\,m$^2$ of liquid scintillator shielded by lead to provide muon detectors with an energy threshold of 250\,MeV and a muon spectrometer, as described in Sec.\,\ref{sec:Cornell}.  For part of the run-time various configurations of plastic scintillator were operated and were used for a variety of purposes including detector and inter-array comparisons.

The determination of the energy of the primary presented particular problems as it was impossible to relate the observed signal to the number of charged particles, as had been done at Agassiz, in a reliable way.  Measurements made with muon, scintillator and water-Cherenkov detectors at the same point in the shower, showed that, on average for every electron there were roughly 10 photons \citep{Kellermann:1970ve} and that the mean energy of both electromagnetic components was about 10\,MeV although these numbers changed with distance from the shower core.  The first efforts to go directly from what was observed to the primary energy was to measure the energy deposited in a pool of water envisaged as an annulus of inner and outer radii of 100 and 1000\,m and with a depth of 1.2\,m.  This energy deposit, $E_{100}$, was related to primary energy using model calculations developed by Hillas and A J Baxter in the late 1960s.  Empirical descriptions of the hadronic features of showers were combined with calculations of the electromagnetic cascades in early Monte Carlo calculations carried out on a mainframe computer with only 64\,kByte of memory.  At 500\,m from the axis of an event produced by a primary particle of $\sim 10^{18}$\,eV about half of the energy flow is carried by electrons and photons 
that are very largely absorbed in the water with the remaining energy being carried by muons.  It was found the $E_{100}$ was $\sim 1/160$th of the primary energy \citep{Baxter:1969vh}.  This method of estimating the primary energy was used to argue that a shower of energy $> 5 \cdot 10^{19}$\,eV had been detected at Haverah Park soon after the Greisen-Zatsepin-Kuzmin prediction (1966) that such events should be rare \citep{Andrews:1968up}.

Two problems with this approach were soon identified: firstly, in large events there was rarely a detector close to 100\,m from the shower axis and secondly, lack of  knowledge of the variation of the lateral distribution function with energy led to a systematic uncertainty that was hard to evaluate.  The problem of what parameter to measure was solved in 1969 through a seminal insight of Hillas \citep{Hillas-70} and his solution has been widely adopted in subsequent measurements with ground arrays, at AGASA, the Auger Observatory and the Telescope Array.  Hillas, using a sample of 50 Haverah Park events, showed that if $E_{100}$ was found using different power laws to describe the lateral distribution function the differences in the derived values was large, $\sim 1.7$.  It is worth quoting Hillas's insight directly from his paper

\begin{quote}
{\it However, because of the geometry of the array, the alternative fits to the data obtained with the different structure functions [now called lateral distribution functions] are found usually to give the same answer for the density $\rho_{500}$ at 500\,m from the axis: in the sample examined $\rho_{500}$ was usually altered by less than 12\,\% by the different assumptions.  A similar effect will arise in other large arrays, the exact distance $R$ for which the density is well-determined depending on the detector spacing.}
\end{quote}

For the larger Haverah Park array, which was being brought into operation at the time of Hillas's insight, the parameter $\rho_{600}$ was adopted with $S(600)$ and $S(1000)$ being chosen for the AGASA and Auger ground arrays respectively, where $S$ denotes the signal at the appropriate distance in scintillators and in water-Cherenkov detectors of the same 1.2\,m depth as those used at Haverah Park.

Of course model calculations are still required to obtain the primary energy from $S(r_{\rm opt})$.  Using a set of phenomenologically-based models, Hillas and colleagues showed \citep{Hillas-71} that the fluctuations in $\rho_{500}$ were typically less than 20\,\% for showers produced by proton primaries over a range of energies and furthermore the relationship between $\rho_{500}$ and the primary energy did not depend greatly on the primary mass.  For example for a model that gave a reasonable fit to several features of the Haverah Park data, the difference between the energy estimated assuming a proton primary $A = 1$ and one of $A = 10$ was below 10\,\% for a range of energies of 100.  These calculations were the basis of the energy spectrum reported by the Haverah Park group in 1990 \citep{Lawrence-91}.  The spectrum was revised 10 years later with a more modern model of hadronic interactions \citep{ave01}: it was found that the energies estimated for the largest events had to be reduced by $\sim 30$\,\%.  Some of the difference was attributable to a better simulation of the response of the water-Cherenkov detectors.

In the early 1970s, following the promptings of Greisen, air-Cherenkov light was measured using 5'' PMTs pointing upwards without mirrors by K E Turver and his colleagues using showers produced by primaries of energy $> 10^{17}$\,eV detected with the array of water-Cherenkov detectors at Haverah Park.  The potential of using this radiation to cross-calibrate estimates of shower energies made at different arrays was explored by Turver and K J Orford who proposed transporting a single-PMT detector from array to array to measure the signal at a reference distance. Attempts to carry out a cross-calibration between Volcano Ranch and Haverah Park in 1974 failed for a variety of reasons.  This idea might still be useful to help understand remaining discrepancies between the different sets of observations.

Advantage was taken of the large area (34\,m$^2$) of the four central detectors of the array to study the thickness of the shower front in some detail.  Using the rise time of the signals in these detectors Watson and Wilson \citep{Watson:1974wq} were able to demonstrate for the first time that there were measurable differences between showers implying that large fluctuations in shower development were detectable.  Using Linsley's ideas of elongation rate, the rise time were also used to infer the rate at which the depth of shower maximum changed with energy and the fluctuation of it, without recourse to model calculations \citep{Walker-81,Walker-82}.

For a period of 4 years, about 0.5\,km$^2$ of the array was enhanced with $30 \times 1$\,m$^2$ water-Cherenkov detectors with 150\,m spacing.  As at Volcano Ranch, these additional detectors allowed precise measurements of the lateral distribution to be made. Within limits set by hadronic interaction models the fraction of protons at $2 \cdot 10^{17}$-$10^{18}$\,eV was found $(34 \pm 2)$\,\% \citep{ave01}.

Thirteen years after the Haverah Park project was closed, an effort to develop techniques for analysing the showers with large zenith angles that had been recorded there was begun with the aim of understanding the background against which any showers initiated by neutrinos would need to be picked out in studies that were foreseen at the Auger Observatory \citep{Capelle-88}.  Inclined showers were known to lose circular symmetry because of the deflection of particles, dominantly muons, in the geomagnetic field at angles above $60^\circ$ \citep{Andrews-71} and it was not possible to study such events in much detail because of limitations in the computing power then available.  This study \citep{Ave-2000a} led to a method of inferring the primary energy by matching maps of the distribution of the muon signals against what was predicted by shower models was devised, a technique has subsequently been adopted in detailed studies of inclined events recorded at the Auger Observatory, and to a limit to the flux of photons above $10^{19}$\,eV \citep{Ave-2000b} which was sufficiently strong to rule out some of the models of the photon fluxes expected from super-heavy relic particles then popular.

The work of Allan and his colleagues on the radio-emission associated with showers carried out at Haverah Park will be discussed below. The site was also used as a test-bed for some ideas, including the use of GPS receivers to obtain timing information, adopted later at the Auger Observatory.

\subsection{The Yakutsk Array}
\label{sec:Yakutsk}

By far the most complex, and most northerly, of the early giant arrays, was that operated by the Institute of Cosmophysical Research and Aeronomy at Yakutsk, Siberia (105\,m).  This array, originally proposed by D D Krasilnikov in 1959, began taking data with 13 stations in 1967 and was developed to cover an area of 18\,km$^2$ in 1974. The leaders were D D Krasilnikov and N N Efimov with the close involvement of Nikolsky and Khristiansen from Moscow.  A detailed description of the array can be found in \citep{Afanasiev-93}. A particularly important feature was the presence of 35 PMT systems of various areas to measure the air-Cherenkov radiation associated with the showers. These gave indirect information about the longitudinal development of showers and provided a calorimetric approach to the energy estimates for the primary particles through the track-integral method. Measurements relating to the energy spectrum, the mass composition and arrival direction distribution of cosmic rays above $10^{17}$\,eV have been reported.  In recent years the array has been contracted to study showers of lower energy and to make more detailed investigations of showers of higher energy.

\subsection{The SUGAR Array}
\label{sec:Sugar}

The team from the University of Sydney who designed `The Sydney University Giant Air Shower Recorder (SUGAR)' introduced a totally novel concept to the detection of extensive air showers by an array of ground detectors.  Before this innovation, the practice had been to link the detectors with cables to some common point where coincident triggers between them could be made and the signals recorded, in the early days often using oscilloscopes.  This method becomes impractical for areas much above 10\,km$^2$ as it was rarely possible to have the relatively unrestricted land access enjoyed by Linsley at Volcano Ranch: the cost of cable, their susceptibility to damage and the problems of generating fast signals over many kilometres were further handicaps.  The concept, due to Murray Winn, was first discussed in \citep{McCusker-63}.  The Sydney group proposed the construction of an array of detectors that ran autonomously with the time at which a trigger above a certain number of particles was recorded being measured with respect to a timing signal transmitted across the area covered by the detectors.

The concept was realised in the Pilliga State Forest near Narribri (250\,m)
where 47 stations were deployed over an area of $\sim 70$\,km$^2$.  Most of the detectors were on a grid of 1\,mile (1.6\,km) with 9 on a smaller spacing to enable smaller showers to be studied.  Time and amplitude data were recorded locally on magnetic tape and coincidences between different stations found off-line some time after the event.  A difficulty was that the rate of triggers of a local station above a level that was low enough to be useful is very high and the rate could not be handled with technologies available at the time.  The problem was solved by burying the detectors under 2\,m of earth and placing them in pairs 50\,m apart.

While the concept was brilliant it was somewhat ahead of its time in terms of the technology available.  Calor gas had to be used to supply the power at each station and the reel-to-reel tape recorders proved difficult to maintain in the dusty environment.  The problem of handling many magnetic tapes at a single computing site proved to be a challenge.  The PMTs used were 7'' in diameter and suffered from after-pulsing which complicated the measurement of the signals as logarithmic time-to-height converters were used to find the amplitudes \citep{Suga-61}.  Efforts were made to overcome this difficulty.  There was also a serious problem in estimating the energy of events as only muons were detected and therefore there was total reliance on shower models with little ability to test which was the best to use because of a lack of different types of detector in the array.  Attempts to overcome this with a fluorescence-light detector and with a small number of unshielded scintillators were unsuccessful.   Energy spectra were reported in \citep{Winn:1986ua}.

The measurement of the shower directions to a precision of a few degrees was a demonstration that the timing stamp method was effective and the most valuable data from the SUGAR array were undoubtedly from the measurements of directions, the first such measurement to be made from the Southern Hemisphere at energies above $10^{18}$\,eV \citep{Winn:1986wi}.  No anisotropy was found.  In later analyses of the SUGAR database, the Adelaide group reported the detection of a signal from the region of the Galactic Centre \citep{clay00,bellido01}.

The concept of autonomous detection was tested at Haverah Park in an early attempt to devise methods to construct an array of $\sim 1000$\,km$^2$ \citep{Brooke-83} but the method had its most effective realisation in the system that was designed for the surface detector array of the Auger Observatory and subsequently at the Telescope Array.

Catalogues of showers recorded at the Volcano Ranch, Haverah Park, Yakutsk and SUGAR arrays are available \citep{Wada-86}.

\subsection{Akeno and AGASA}
\label{sec:AGASA}

The largest shower array constructed before the advent of the Auger Observatory and the Telescope Array was the `Akeno Giant Air Shower Array' (AGASA) which was built in Japan outside Tokyo at Akeno (900\,m).
The AGASA team was led by Nagano and the array operated from 1990 until 2004.  It consisted of 111 unshielded scintillator detectors each of 2.2\,m$^2$ with an inter-detector spacing of $\sim 1$\,km.  Muon detectors of various areas between 2.4 and 10\,m$^2$ were installed at 27 of the 111 detectors.  Each detector was serviced using a detector control unit that recorded the arrival time and size of every incident signal and logged monitoring information, the pulse height distribution, the voltage, counting rate and temperature in a manner that anticipated what is done at the Auger Observatory.  An optical fibre network was used to send commands, clock pulses and timer frames from the central station to each module and to accept the trigger signals, shower data and monitoring data.

An area of 1\,km$^2$, the Akeno array, devised by K Kamata as a prototype for the giant 100\,km$^2$ array, was covered with a denser array of scintillators and was brought into operation in 1979.  This array became part of the larger complex but was also used to measure the energy spectrum \citep{nagano84a} and the $p$-air inelastic cross-section \citep{Hara:1983vg,Honda:1993vc}.

Exploratory work for future arrays was also carried out at AGASA.  Three detectors, with two scintillators sandwiching a 1\,cm thick lead plate, were used to explore the electron, photon and muon component far from the shower core \citep{Honda-87} and two prototype water-Cherenkov detectors, as planned for the Auger Observatory, were tested \citep{Sakaki-97}.

Some important claims were made about the energy spectrum and the arrival direction distributions at the highest energies.  The energy spectrum was reported as extending beyond $10^{20}$\,eV with the 11 events observed, showing no sign of any cut-off.  The energies were estimated using model calculations and subsequent work, in which the energy spectrum has been found by the track-length integral method inferred from observations of fluorescence light, have shown that there were deficiencies in the model calculations used.

\section{Use of the Monte Carlo Technique}
\label{sec:MC}

The use of Monte Carlo techniques in the study of the cascade characteristics of air showers has grown enormously since they were first introduced in the early 1960s.  The techniques developed have become indispensable for the interpretation of data, to model the performance of detectors and to understand the development of the cascade itself.  Two examples of the application of the technique have already been given in the discussions of fluctuations in electromagnetic cascades (c.f. Sec\,\ref{sec:fluctuations}) and in the sharing of energy between the constituents of helium nuclei in atmospheric collisions (c.f.\ Sec\,\ref{sec:Cores}).  Wilson's work \citep{Wilson:1952wc} was carried out with what was essentially a roulette wheel but subsequent activities depended on the computing power available with particular ingenuity being shown in the earliest days to combat the limitations of the times.

Early calculations of the cascade development made use of phenomenological models of the hadronic interactions such as the CKP-model of Cocconi, Koester and Perkins \citep{Cocconi-62} developed to calculate particle fluxes at future accelerators.  Other phenomenological models were developed and were used in interpretation of data from many experiments.  A problem was recognised by Linsley in 1977 when he found that some of the Monte Carlo calculations produced results that were in violation of his elongation rate theorem \citep{Linsley-77-elong} (c.f.\ Sec.\,\ref{sec:Volcano}) in that the computation of the change of some shower parameters with energy was greater than was physically possible.  This raised questions about the accuracy of some of the Monte Carlo codes. Accordingly Linsley and Hillas \citep{Linsley-82}
organised a discussion targeted at having interested groups use a common model within their codes to calculate the depth of shower maximum and how it varied with energy.  This exercise was partially successful and the results from seven groups who contributed were reported and assessed.  From an analysis carried out on the individual reports it was shown that one cascade algorithm did not conserve energy.  In addition a number of issues were identified such as the use of different values for the radiation length in different codes, the importance of energy conservation, the importance of pion decay and the effect of limiting the energy below which particles are not followed.  The issue of the radiation length to be adopted was subsequently considered further by Linsley \citep{Linsley-85} who recommended a value of 37.15\,\gcm.  The inclusion of pion decay and the limitations to the energy that a particle could be tracked resulted from restrictions set by the computer power available in different institutions.  The problem of following all of the particles in a shower was also discussed by Hillas \citep{Hillas:1981ua}: he introduced the concept of `thinning' which has subsequently had very wide application.  He pointed out that it was not necessary in some cases to follow every particle to get a good picture of a shower and reported that good results for muons were obtained by choosing a demarcation energy, $D$, set at $10^{-4}$ of the primary energy, and following all particles of energy $> D$ but only a fraction of particles of energy $E < D$.  The technique was also used for electromagnetic cascades.


By the mid-1980s computing power had increased enormously and several major programs were developed.  Hillas created the MOCCA program at this time, written in Pascal.  Only a limited description of this code reached the literature but it was made available to the designers of the Auger Observatory for which purpose it was translated into FORTRAN in the early 1990s.  The most readily accessible account of the thinning method and of MOCCA is available in \citep{Hillas-97}.

When work on the KASCADE project at Karlsruhe (c.f.\ Sec.\,\ref{sec:KASCADE}) started, it had been realized that most of the cosmic ray projects used their own specific tools for EAS simulation to infer astrophysical quantities. Thus, in the not-infrequent cases of disagreement between experiments it remained unknown whether the problem had been of purely experimental nature related to the apparatus or whether it had been due to differences in the EAS simulations applied. Thus, parallel to preparing for constructing KASCADE, an extremely important code was developed, with input by J Capdevielle and by P Grieder who was one of the early pioneers of the Monte Carlo method and who, in Bern, had access to exceptionally powerful machines.  The CORSIKA code (`COsmic Ray SImulation for KAscade'), continuously maintained by a team at Karlsruhe with support from all over the world, has the merit of allowing different models of nuclear interactions to be included in an easy way and the authors made it widely available to the community. Thus, over the years it had become a {\it de facto\/} standard in the field, similar to the GEANT simulation package in high-energy physics. CORSIKA allows the study of the evolution of EAS in the atmosphere initiated by photons, protons, nuclei, or any other particle and its applications range from classical EAS arrays to Cherenkov telescope experiments covering the energy range from $10^{12}$\,eV up to the highest energies observed ($E_0 > 10^{20}$\,eV ). The important step made with CORSIKA is that even though the EAS modeling may not be perfect, the very same modeling is used by all major experiments in the field.  J Knapp and D Heck were two of the major drivers behind the CORSIKA project. As Knapp pointed out in his ICRC rapporteur talk \citep{Knapp-97}:
\begin{quote}
{\it Is the composition changing or not? The answer depends on the yardstick (i.e.\ the Monte Carlo program) used for comparison. Use the same yardstick to get consistent results, use a well calibrated yardstick to get the correct result.}
\end{quote}


In addition to its application in shower modelling, the CORSIKA code has been used in many other investigations, ranging from mountain and pyramid tomography trough muon measurements over neutrino searches to the possible link between cosmic rays and climate (see e.g.\ \citep{Usoskin:2006kb}).
Another advantage of intense use by many groups is that programming errors or conceptual faults are quickly identified and eliminated. Nevertheless, in the context of the Pierre Auger Observatory another code, named AIRES\footnote{S.J. Sciutto, AIRES User’s Manual and Reference Guide; available electronically at www2.fisica.unlp.edu.ar/auger/aires}, had been developed in La Plata and is very useful for cross check studies and specific simulations.


It should be noted that modern codes have become so complicated that sanity has been maintained in some instances by recalling the simplified description of electromagnetic showers given by Bethe and Heitler (1934) and of hadronic cascades by Matthews (2005).

\section{The Impact of the Discovery of the Microwave Background Radiation}
\label{sec:GZK}

The primary purpose of the early km$^2$-scale EAS experiments was to study the energy spectrum and arrival directions of ultra-high energy primary cosmic rays for the information which these data give about the origin of cosmic rays. It had been realised that cosmic ray particles beyond $10^{20}$\,eV, which were believed to be atomic nuclei, would have a very great magnetic rigidity. Thus, the region in which such a particle originates must be large enough and possess a strong enough magnetic field so that $RB \gg (1/300) \cdot (E/Z)$, where $R$ is the radius of the region (cm), $B$ is the magnetic field in Gauss, and $E$ is units of TeV. Also, anisotropies were expected to be seen. However, estimates of the particle flux were over-optimistic.

\begin{figure}[t]
\centerline{\resizebox{0.7\textwidth}{!}
{\includegraphics{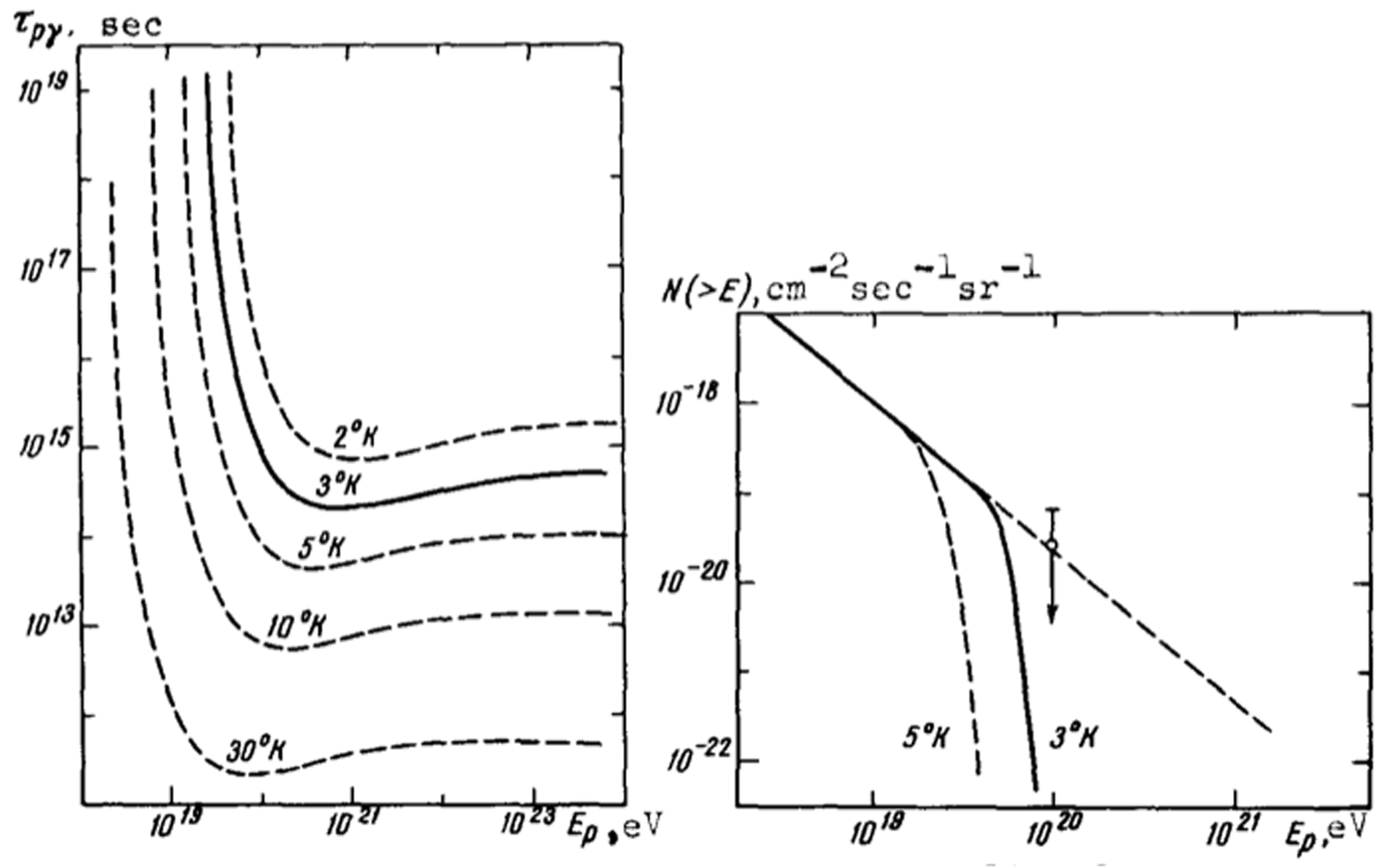}}}
\caption{
Left: Characteristic time for GZK-like collisions as a function of proton energy for different photon gas temperatures. Right: Expected suppression of the energy spectrum for a simplified source scenario \citep{Zatsepin:1966jv}.}
\label{fig:Zatsepin-GZK}
\end{figure}

In May 1965 Penzias and Wilson reported their serendipitous observation of the cosmic microwave background radiation (CMB) \citep{Penzias-95}. Only a few months later, Gould and Schr\'{e}der \citep{Gould:1966tu} pointed out that high-energy photons of a few $10^{14}$\,eV traversing cosmic distances would suffer rapid energy losses due to electron-positron pair production by photon-photon collisions in the CMB. Thus, some earlier claims of high-energy muon-poor showers, supposed to be initiated by photons of extragalactic origin, were questioned by the authors and no ``window'' was open for extragalactic $\gamma$-ray astronomy until well-above $10^{14}$\,eV \citep{Jelley:1966ts}. A few months later, Greisen \citep{Greisen-66b} and independently Zatsepin and Kuz'min \citep{Zatsepin:1966jv} noted a related effect for proton primaries in the CMB, in this case photo-pion production being responsible for rapid attenuation of protons of energy beyond $4\cdot10^{19}$\,eV. Figure \ref{fig:Zatsepin-GZK} shows the key figure of Zatsepin \& Kuz'min's paper including the data point from \citep{Linsley:1963uo} which was hard to understand after this finding. The title of Greisen's paper ``End to the Cosmic-Ray Spectrum?'' expressed the situation perfectly and the effect became known as {\em ``GZK-effect''}. Its worth pointing out that both Greisen as well a Zatsepin \& Kuz'min also noted that light and heavy nuclei would suffer rapid photo-disintegration above about the same energy threshold.

It is an interesting fact that the large shower arrays that were developed in the UK, Siberia, and in Australia which dominated the studies of cosmic rays above $10^{17}$\,eV in the 1970s and 1980s were all planned before the discovery which was to become one of the main motivations for their operation.  By contrast planning of the Fly's Eye detector, which detected fluorescence radiation, was begun in 1973 long after the GZK-effect had been recognised and its verification became one of the prime motivations for its construction.  However it turned out that none of these devices had a sufficiently large aperture to establish the existence of a steepening in the cosmic ray spectrum. In fact, the dispute between AGASA and Fly's Eye about the observation of a suppression at the highest energies became an important argument for the construction of Pierre Auger Observatory by the end of the 1990s (see next sections).

\section{The Development of the Fluorescence Technique}
\label{sec:Fluorescence}

A powerful new technique for studying extensive air showers from the highest energy particles was developed during the 1960s.  The approach depends on observing the faint fluorescence radiation which is emitted isotropically when the 2P and 1N band-spectra associated with molecular nitrogen are excited by ionising particles.  It allows the atmosphere to act as a massive calorimeter and in principle gives the possibility of measuring the energy of cosmic rays without resorting to assumptions about hadronic physics.  The key to large apertures is the isotropic emission, with application of the track-length concept used to give the shower energy.

It is not clear who first had the inspiration of using the excitation of N$_2$ for cosmic ray work and it may well be that the idea occurred to several people at about the same time.  The concept of employing air-fluorescence to detect $X$-rays from nuclear explosions appears to have been discussed post-the Manhattan project
and it seems probable that Edward Teller was the first to have the idea of using air-fluorescence induced by $X$-rays produced in such explosions as a monitoring tool. This is the so-called ``Teller light''. The documents are still classified, but the application to NSF made by the Utah group in 1973 (J W Keuffel et al.) for construction of the Fly's Eye detector refers to the Teller light in the title of a classified paper. Greisen, who had been at the Trinity test, was perhaps aware of this activity and the idea may have been discussed informally in the US during the 1950s as a way of detecting the highest-energy cosmic rays\footnote{S Colgate, private communication to AAW}.

The method was first discussed at an international forum in La Paz in 1962 where Suga outlined the idea and showed a spectrum of the emission in the ultra-violet part of the spectrum using $\alpha$-particle sources \citep{Suga-62}.  The signal was expected to be small, even from showers produced by primary cosmic rays of $10^{20}$\,eV, as the isotropic emission is only about 4 photons per metre of electron track in the wavelength range from 300 to 450\,nm.  

\begin{figure}[t]
\centerline{\resizebox{0.5\textwidth}{!}{\includegraphics{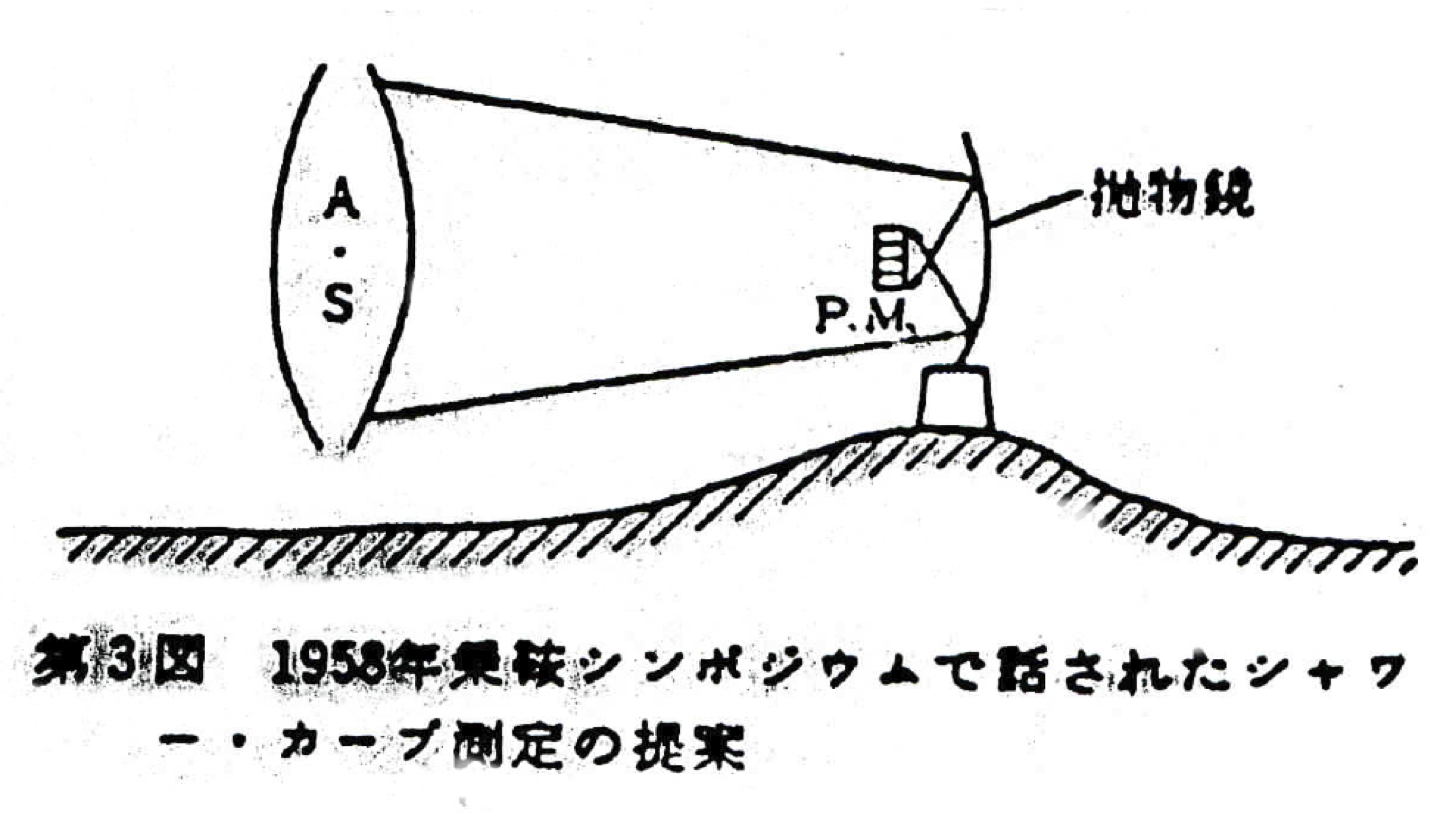}}}
\caption[xx]{
First documented concept of a PMT camera viewing the fluorescence light from an air shower collected with a mirror.  The similarity of the layout shown here to the devices constructed by the Utah, Auger and TA groups is remarkable. From Proceedings of Norikura meeting in summer 1957 (INS Report 1958).}
\label{fig:Fluorescence-Jp}
\end{figure}

The fact that the light is emitted isotropically makes it feasible to observe showers `side-on' from very great distances and thus it opens the possibility of monitoring large volumes of air.  It is clear from a diagram taken from a Japanese publication of 1958 (Fig.\,\ref{fig:Fluorescence-Jp}) that discussions about using this method to detect high energy cosmic rays must have taken place in Japan, under the guidance of Suga for some years prior to his report at La Paz\footnote{Tanahashi and Nagano, private communication to AAW}.  During the discussions following Suga's presentation, Chudakov reported the results of measurements that he had made in 1955 - 1957 of the same phenomenon.  He examined this effect as he was concerned that it might be a background problem in the detection of Cherenkov radiation, a technique that was being developed strongly in the Soviet Union in the 1950s, but he was slow to write up his observations \citep{Belyaev-1966}. Chudakov also observed transition radiation in the same series of experiments.

The use of fluorescence radiation to detect air showers was 
studied in Greisen's group which included Alan Bunner \citep{Bunner-67,Bunner:1968wo} who measured the spectrum of the light produced by particles in air.  Greisen did not mention this activity at La Paz but in an important review talk in 1965 \citep{Greisen-66b} he pointed out many of the key issues and showed the band spectrum of the fluorescence light from 200-460\,nm.  This paper had a much wider distribution than did the report of Suga's talk in La Paz.

It is unclear whether Suga and Oda's input was a further independent effort or whether Oda had picked up the idea while he was at MIT.  Rossi would probably have been aware of the possibilities of $X$-ray detection from nuclear explosions by the fluorescence technique as he too had worked on the Manhattan project but Teller's work was classified.

The Japanese plans did not develop immediately.  Tanahashi from the INS group worked in Greisen's group at Cornell in the mid-1960s where efforts were being made to detect fluorescence radiation using a set of Fresnel lenses.  On his return to Japan Tanahashi played a major role in setting up a fluorescence detector at Mt Dodaira, with Fresnel lenses, and the successful detection of air showers by the fluorescence method was reported in 1969 \citep{Hara-1970}.  Greisen acknowledged this achievement generously\footnote{Letter from Greisen to Tanahashi, 29.\ Sept.\ 1969} and recently Bruce Dawson has confirmed the INS conclusions using his experience from the Auger Observatory to re-examine the INS data \citep{Dawson-2011}. The use of fluorescence light as a detection technique seems to have been thought of more or less simultaneously in three countries but it is clear that the Japanese air shower physicists were the first to make convincing detections.

The work of Greisen's group at Cornell ended in 1972.  Although unsuccessful it is worth quoting the closing paragraphs of his final report to the US Atomic Energy Commission:

\begin{quote}
{\it It became apparent that the detection of air showers by fluorescent light could only be made successfully by (a) operating in a different part of the earth where the weather would permit observing during four times as many hours per year, and where the lower atmosphere is free of the particles and aerosols that cause Mie scattering; and (b) taking full advantage of modern electronic technology in the information processing, so as to separate the air shower patterns from the background noise without loss or degradations of information in doing so.  This would be an engineering task of considerable magnitude and cost. \ldots With considerable relief at the termination of a long period of arduous and rather unrewarding effort the recording stations were shut down in January 1972 ten years after initiating the proposal that the work be begun.}
\end{quote}

Despite this down-beat coda, Greisen's efforts had inspired many.  Tanahashi attempted to introduce the fluorescence technique into the Sydney Air Shower array and Greisen's work was taken up in the USA by a team at the University of Utah, led first by Jack Keuffel.
Following the Japanese efforts, another convincing demonstration of the method was finally achieved through the operation of a small fluorescence detector in coincidence with the Volcano Ranch scintillator array \citep{Bergeson:1977uc}.  Fluorescence detectors could now be used as stand-alone devices.

\begin{figure}[t]
\centerline{\resizebox{0.4\textwidth}{!}{\includegraphics{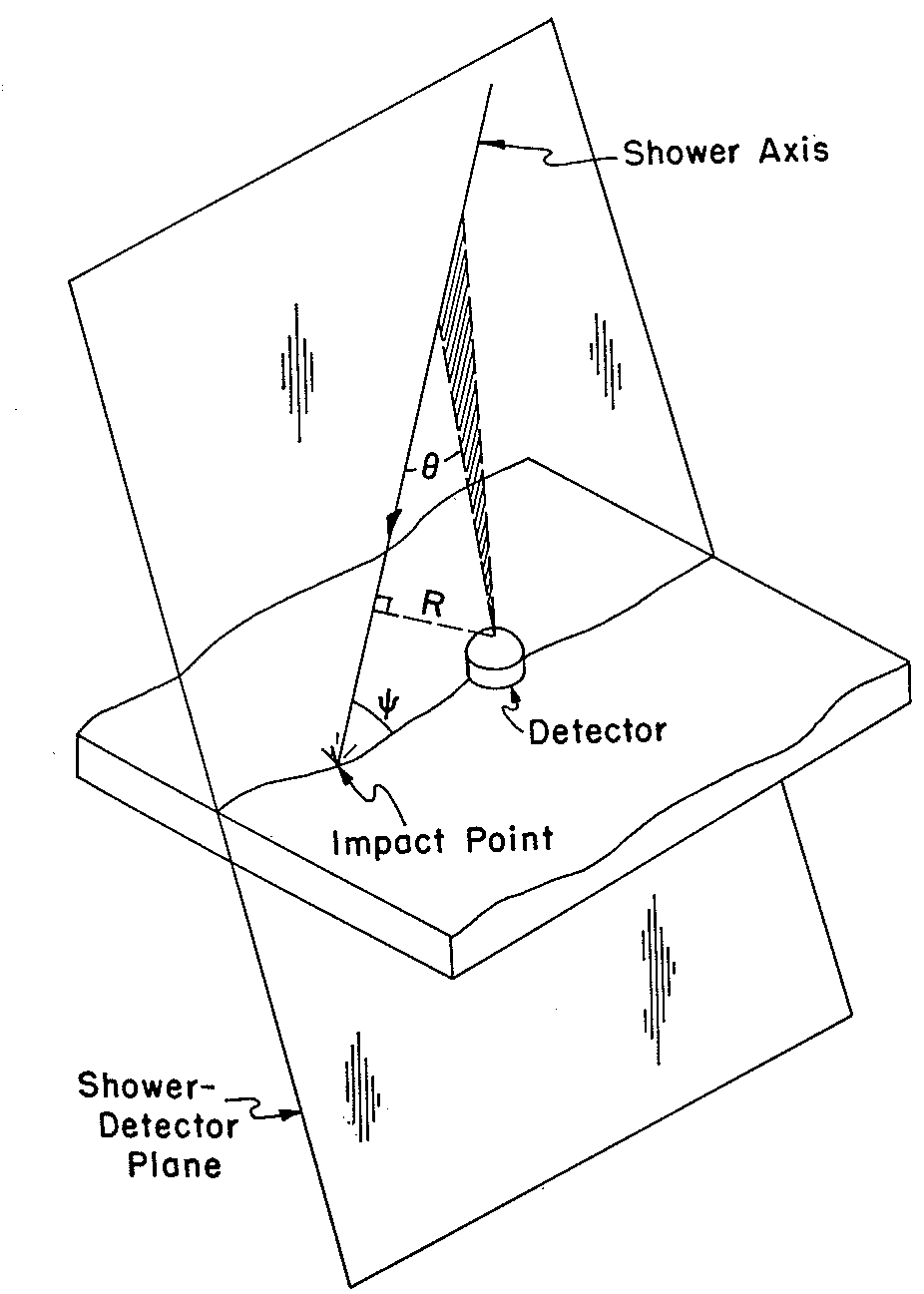}}}
\caption{Perspective view of the shower geometry for fluorescence detector observations \citep{Bunner-64}.
}
\label{fig:shower-plane}
\end{figure}

A lasting legacy of Greisen's work was a sketch made by Bunner for his 1964 master's thesis (Fig.\,\ref{fig:shower-plane}).  Here the essence of the reconstruction method is shown: the diagram has been reproduced many times but its source has rarely been acknowledged.

\section{Development of Fly's Eye}
\label{sec:FlysEye}

Following the death of Keuffel, the Fly's Eye efforts in Utah were led by Haven Bergeson and then by George Cassiday and Gene Loh. Construction of a prototype detector (which later became known as Fly's Eye I) consisting of 67 camera units started near Dugway (Utah) in the early 1970s \citep{Bergeson-1975} and for cross-correlation and overall testing, three of those were taken to the Volcano Ranch array and positioned about 1.5\,km from the ground array. With this set-up 44 showers were registered in 12 consecutive nights of operation. The events recorded ranged up to $2.5 \cdot 10^{18}$\,eV and established the method of air fluorescence detection \citep{Bergeson:1977uc} marking a major breakthrough in ultra high-energy cosmic ray detection methods.

The 67 units of the Fly's Eye I detector consisted of 1.5\,m diameter front aluminized spherical section mirrors, associated Winston light collectors, PMTs and data acquisition electronics. The Winston light collectors and PMTs were hexagonally packed in groups of either 12 or 14 light sensing ``eyes'' mounted in the focal plane of each mirror. A motorised shutter system kept the eyes both light tight and weather proof during the day and permitted exposure to the sky at night. Each mirror unit (and associated light sensing cluster) was housed in a single, motorised corrugated steel pipe about 2.13\,m long and 2.44\,m in diameter giving the Fly's Eye detector a very specific look. In total, there were 880 PMTs at Fly's Eye I, each subtending a $5^\circ \times 5^\circ$ pixel of the sky, which completely imaged the entire night sky. 

To improve shower reconstruction in the absence of a ground array at Dugway, the stereoscopic observations were pioneered by erecting Fly's Eye II at 3.4\,km distance relative to Fly's Eye I. This smaller array of 8 identical units (later extended to 36) started operation in 1986. Fly's Eye II observed roughly one azimuthal quadrant of the night sky with elevation angles ranging between $2^\circ$ and $38^\circ$ above the horizon. In monocular mode, Fly's Eye reached a collection area of about 1000\,km$^2$ (effectively about 100\,km$^2$ if the $\sim 10$\,\% duty cycle of night time operation is taken into account).

A spectrum from a single eye was reported in 1975 along with a measurement of the mass composition above $10^{18}$\,eV before work with two Eyes started.
The science output culminated in the report of an event of $(3.2 \pm 0.9) \cdot 10^{20}$\,eV (51 Joule) recorded in 1991 \citep{bird95}, still the highest energy ever claimed.
The event fell only 12 km from the Fly's Eye I detector allowing a good measurement of its profile and energy. However, it fell behind the Fly's Eye II detector so was not seen in stereo.

The aperture of this pioneering experiment was too small to measure the spectrum at $10^{20}$\,eV and hence to observe the GZK cutoff. However, the Fly's Eye and AGASA spectral measurements (see below) set the stage for work to come with the HiRes and the Pierre Auger Observatories.

\section{The Cygnus X-3 Story and its Impact}
\label{sec:Cygnus}

One of the consequences of the work at the Kiel array was the impact of an unexpected result that was never confirmed.  After detailed studies on the cores of showers had been completed, attempts were made to search for point sources of cosmic rays using the excellent angular resolution, $\sim 1^\circ$, of the array.  In 1983 Samorski and Stamm \citep{Samorski:1983vo} reported a surprising observation suggesting that the 11\,kpc distant $X$-ray binary system, Cygnus X-3, was a source of photons of above $2 \cdot 10^{15}$\,eV.  A signal of $4.4 \sigma$ was found in the region around the object using data obtained between 1976 and 1980 based on 16.6 events above a background of $14.4 \pm 0.4$.  Cygnus X-3 has a periodicity of 4.8 hours and 13 of the events in the on-source region were in one of the 10 phase bins into which the 4.8 hour period was divided.  The Kiel conclusion appeared to be confirmed by results from a sub-array at Haverah Park \citep{LloydEvans:1983ue}, tuned to $\sim 10^{15}$\,eV, and also by measurements made around the same time at lower energies using the air-Cherenkov technique.  The claims stimulated great interest and, although now regarded as incorrect, gave a huge stimulus to activity in the fields of high-energy $\gamma$-ray astronomy and ultra-high energy cosmic rays.

For the air shower field an important consequence was the interest that James W Cronin (University of Chicago) took in the subject.  A Nobel Laureate for his work in particle physics, Cronin entered the cosmic ray field with vigour and led a team from the Universities of Chicago and Michigan to construct an air shower array, known as CASA-MIA, of $\sim 0.24$\,km$^2$, to search specifically for signals from Cygnus X-3 \citep{Borione:1994wt}.  The array was on a different scale, in terms of numbers of detectors, from anything built previously with 1024 scintillators of 1.5\,m$^2$ laid out on a rectangular grid with 15\,m spacing, above the muon detectors, each of 64\,m$^2$, buried 3\,m deep at 16 locations.  As with the Chacaltaya array built 30 years earlier, the idea was that showers with small muon numbers were likely to be produced by $\gamma$-rays. The area of the muon detector was over 40 times that at Chacaltaya.

No signals were detected from Cygnus X-3 suggesting that the results from Kiel, Haverah Park and the TeV $\gamma$-ray observatories were spurious.  However, what this enterprise showed was that it was possible to build much larger detectors than had been conceived previously and Cronin went on to be the leading player in the planning and implementation of the Pierre Auger Observatory.  Another consequence of the Cygnus X-3 period was that other particle physicists, most notably Werner Hofmann and Eckart Lorenz, began work at La Palma to search for signals from Cygnus X-3 using a variety of novel methods, but they quickly moved into high-energy $\gamma$-ray astronomy.

\section{Recent and Current Activities}
\label{sec:Current}

\subsection{Some More Recent Projects}

The Cygnus X-3 observations revitalized experimental efforts to study cosmic rays above $10^{14}$\,eV which resulted in a new generation of devices with sophisticated instrumentation, including CASA-MIA, GRAPES, HEGRA, EAS-TOP, KASCADE, MAKET-ANI, Tibet-AS$\gamma$, and the SPASE array at the South Pole. Obviously, pointing resolution had been an important design criterion. Searching for primary photons by EAS observations also required at least limited sensitivity to the primary composition, either by adding muon counters or air-Cherenkov detectors. Below, we shall briefly describe a few of those, highlighting their major results. None of them could confirm the observations by the Kiel array, but the results of these experiments - thanks to their multi-purpose instrumentation -  advanced our understanding of high-energy cosmic rays rigorously and shaped a new community. 

\subsubsection{EAS-TOP}

In Italy a group led by Gianni Navarra in the mid 1980s started to install a multi-component detector at the Campo Imperatore at 2005\,m a.s.l.\ on top of the underground Gran Sasso Laboratory. 
Optimised under the design criteria laid out, EAS-TOP started operation in 1989 and consisted in its final stage of an array of 35 modules of unshielded scintillators, 10\,m$^2$ each, separated by 17\,m near the centre, and by 80\,m at the edges of the array, covering an area of about 0.1\,km$^2$ for measurement of the shower size. A central 140\,m$^2$ calorimeter of iron and lead, read out by 8 layers of positional sensitive plastic streamer tubes, allowed measurements of hadrons ($E_{\rm h} \ge 30$\,GeV) and muons ($E_\mu \ge 1$\,GeV) in the shower core \citep{Aglietta-89}. A very important feature of EAS-TOP was the unique possibility of correlated measurements with the MACRO detector located underground in the the Gran Sasso Laboratory, combining shower information at ground with TeV energy muons measured underground.

The physics portfolio was very rich covering (i) cosmic composition measurements across the knee, demonstrating an increasingly heavier composition towards higher energies. Moreover, by making use of the multi-component measurements it allowed (ii) tests of hadronic interaction models, (iii) measurements of the $p$-Air interaction cross-section, and very importantly, it (iv) provided stringent tests of the cosmic ray anisotropy as a check of decreasing Galactic content of the cosmic rays and providing hints of increasing amplitude and change of phase  above 100 TeV as well as (v) the first observation of the solar Compton-Getting effect.

Before operation was terminated in 2000\footnote{This was primarily for reasons of environmental protection arguments that applied to the Campo Imperatore area that was designated a National Park.}, contacts were made to explore the possibility of shipping the scintillator stations to the KASCADE site in Karlsruhe to continue operation in an enlarged experiment there. A summary of the results from EAS-TOP has been given in \citep{Navarra:2006hp}.

\subsubsection{HEGRA}
\label{sec:HEGRA}

The HEGRA (High Energy Gamma Ray Astronomy) experiment located at the Observatorio de Roque de Los Muchachos (2200 m), La Palma (Canary Islands, Spain) started operation in the mid 1990s. Its science goals were the detection of directed and diffuse cosmic $\gamma$-radiation, the measurement of the chemical composition, and the observation of time-variations in any of the cosmic ray components. For this purpose HEGRA was built as a multi-component detector consisting of on an array of 182 scintillation detectors spread over $180 \times 180$\,m$^2$, a number of Geiger towers, each consisting of six planes of Geiger tubes placed in a support structure of gaseous concrete, and open Cherenkov detectors (AIROBICC) which pioneered the technique of non-imaging Cherenkov observations. Later, the installation was augmented by 6 imaging Cherenkov telescopes which provided the first high resolution measurements of TeV $\gamma$s from the Crab Nebula and from the extragalactic active nuclei Mkn\,501 and Mkn\,421.
The operation of the Geiger counters suffered from gas leakage problems limiting their operation and the prime focus of HEGRA quickly moved to $\gamma$-astronomy before the collaboration essentially split to construct MAGIC and H.E.S.S. Operation of the array was terminated in 2002 by a bushfire which destroyed a large part of the scintillator array. A collection of papers presented at the $28^{\rm th}$ ICRC and covering a range of topics is found in \citep{HEGRA-03}.

\subsubsection{KASCADE and KASCADE-Grande}
\label{sec:KASCADE}
At the end of the 1980s, two institutes at the research centre in Karlsruhe, Germany (now KIT) directed by G Schatz and B Zeitnitz  joined efforts together with University groups from abroad, to construct an EAS experiment at sea level (110\,m) in Karlsruhe. Again, this endeavor and change of research direction away from nuclear physics was motivated largely by the surprising results from the Kiel array, so that $\gamma$-ray astronomy was on the agenda. However, concise measurements the cosmic ray composition and of hadronic interactions were realised to be of great need  and the experiment was designed accordingly. Karlsruhe was chosen as the site mostly because of its direct proximity to all the infrastructure of the research centre needed to operate a most complex EAS experiment. The resulting savings in logistics and operation costs were invested in a very large sampling fraction (actual detector area covered in the fiducial area of the array) of detectors. This was required to compensate for the strong atmospheric overburden affecting measurements in the knee energy range considerably. 

\begin{figure}[t]
\centerline{\resizebox{0.9\textwidth}{!}{\includegraphics{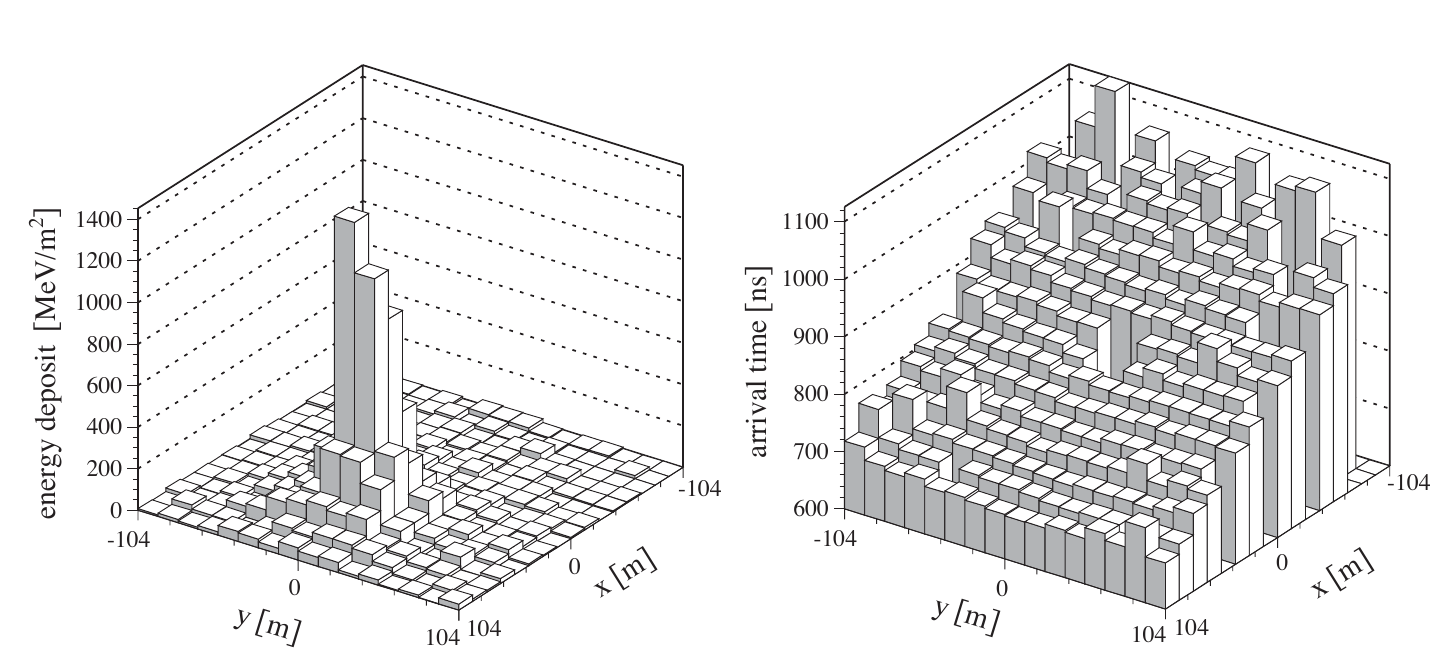}}}
\caption{
Example of an EAS registered by the $e/\gamma$ detectors of the KASCADE experiment in the energy range of the knee. Left: Energy deposits, Right: arrival times. The position of the shower core and the curvature of the shower front are well observed \citep{kascade-03c}.}
\label{fig:eas-in-kascade}
\end{figure}

The KASCADE (KArlsruhe Shower Core and Array DEtector) experiment started data taking in 1996 and consisted of 252 array stations of $e/\gamma$- and $\mu$-detectors spread over $200 \times 200$\,m$^2$, a highly complex 320\,m$^2$ central detector, and a 130\,m$^2$ $\mu$-tracking detector, details of which are described in \citep{kascade-03c}. The sampling fraction of 2.6\,\% and 3.3\,\% for the electromagnetic and muonic component, respectively, is the largest of all EAS experiments ever operated. Like the other projects already mentioned, KASCADE never found any significant diffuse or point-like $\gamma$-flux and only provided upper limits. Its main achievements, however, were tests of hadronic interaction models and most importantly measurements of the cosmic ray composition across the knee. Figure \ref{fig:eas-in-kascade} shows an example event measured with KASCADE. The statistical resolution achieved
in the measurement of the numbers of muons and electrons
above the knee is better than 4\,\% and 10\,\%, respectively. This high precision enabled a two-dimensional unfolding of the measured $N_e$ vs.\ $N_\mu$ distributions -- 45 years after similar plots from the INS array  (c.f.\ Fig.\,\ref{fig:Fukui-nmu-ne}) were analysed \citep{Fukui:1960tl}. The results \citep{KASCADE-05} convincingly demonstrated that the knee in the cosmic ray spectrum is caused by light particles and that a knee could be seen in spectra of five different mass groups with their position shifting towards higher energies with increasing mass, in good agreement with the Peter's cycle \citep{peters61}. This achievement of combining high precision EAS data with sophisticated mathematical tools marked another milestone in cosmic ray physics. 

Obviously this observation showed the need for improved data up to $10^{17}$\,eV where the break of the iron-knee would be expected. The closure of EAS-TOP at about the same time triggered Navarra and KH Kampert and their groups to extend the KASCADE-Experiment with the scintillator stations of EAS-TOP to become KASCADE-Grande \citep{Apel-NIM-10}. It covered an area of about 0.5\,km$^2$ and operated from 2003 to 2010. In a recent paper \citep{Apel:2011bx}, a knee-like structure in the energy spectrum of the heavy component of cosmic rays at $E \simeq 8 \cdot 10^{16}$\,eV was reported. Does this mark the end of the galactic cosmic ray spectrum? In fact, the cosmic energy spectrum appears be much richer in its features than could be described by simple broken power laws, challenges to be addressed by future observations.


\subsubsection{HiRes}

At the highest energies, the second-generation air-fluorescence experiment, High-Resolution Fly's Eye (HiRes) became the successor of Fly's Eye. Proposed in the early 1990s it was completed in 1997 (HiRes\,I) and 1999 (HiRes\,II). It was also located at Dugway, Utah and also had two air-fluorescence detector sites, HiRes I and HiRes II spaced 12.6\,km apart. This detector had smaller phototubes resulting in a pixel size of $1^\circ \times 1^\circ$ in the sky. Amongst other improvements over the original Fly's Eye was an FADC data acquisition system at HiRes II which allowed a much more precise measurement of the longitudinal shower profile.
HiRes took data up to 2006, but the last years of operation suffered from an accident at the military site of Dugway which subsequently meant that a very small number of people could go to the site for shifts.  Despite these operational problems, a rich spectrum of measurements of the cosmic ray composition, $p$-Air cross-section, anisotropies, and the energy spectrum could be reported. Most notably, clear signs of a cut-off in the energy spectrum, in good agreement with the GZK-effect was demonstrated. A comprehensive summary of the late Fly's Eye and early HiRes results can be found in \citep{sokolsky-07}.

\subsection{Current Projects}

As the question about the energy spectrum and elemental composition at the knee has now been largely settled, focus of current EAS experiments has, on the one hand, turned to understanding the putative transition from galactic to extragalactic cosmic rays, believed to occur in the energy range between $10^{17}$ to $5\times 10^{18}$\,eV, and, on the other hand, to the study of the upper end of the energy spectrum. As this articles aims at discussing historical perspectives rather than reviewing the current status of the field, we shall be rather brief here.

The energy range of the knee has mostly been left to the Tibet Array (at Tibet Yangbajing, 4300\,m), IceTop as part of IceCube (located at the South Pole, 3200\,m), GAMMA (on the south side of Mount Aragats in Armenia, 3200\,m), GRAPES (at Ooty, India, 2200\,m), and Tunka (in the Tunka Valley in Buryatia, Siberia, 675\,m). Significant activities are presently seen at the South Pole where construction of IceTop has just been finished and at the Tunka array where open Cherenkov detectors and radio antennae are being installed. The main goal of all of these experiments is to provide new and possibly better information about the composition around the knee energy than has become available from the experiments recently phased-out.

At the highest energies, the Yakutsk array has continued the study of UHECR events started in 1973 with detectors in various configurations. Since 1979, muon detectors with areas up to 36\,m$^2$ (currently, five detectors of 20\,m$^2$ each with threshold energy 1\,GeV for vertical muons) supplement ground-based scintillator stations. The group, however, is rather small and operation in the hostile environmental area difficult.

\subsubsection{Pierre Auger Observatory}

The problem of the small number of events at the highest energies was recognized in the 1980s, even before the AGASA and HiRes detectors had completed construction, and a controversy about the existence or non-existence of a suppression of the cosmic ray flux at the GZK threshold of $5\cdot 10^{19}$\,eV became a major point of discussion. 
This led to the idea that 1000\,km$^2$ of instrumented area was needed if progress was to be made. Cronin argued that 1000\,km$^2$ was insufficiently ambitious and in the summer of 1991 he and Watson decided to try to form a collaboration to build a detector of 5000\,km$^2$, initially without any fluorescence devices. An international workshop, organized in Paris by Murat Boratav in 1992, led to a number of focused studies that culminated in a 6-month Design Study during 1995 hosted at the Fermi National Accelerator Laboratory by the then Director, John Peoples.
Initially named Giant Air shower Project (GAP) it later became the Pierre Auger Observatory, in honour of Pierre Auger's work on the discovery of extensive air showers. 
After the design study two sites of 3000\,km$^2$ in each hemisphere were proposed but this was disfavored by the US funding agencies who would only support a Southern site.
Argentina was selected in a democratic vote at the UNESCO headquarter in Paris in November 1995 and construction of an engineering array finally began in 2001 near Malarg\"{u}e, Argentina. Physics data taking began January 1st 2004 with about 150 water-Cherenkov tanks and 6 fluorescence telescopes and construction of all 1600 surface detector stations covering an area of 3000\,km$^2$ and 24 telescopes finished mid-2008. As of today, Auger has reached an exposure of nearly 25\,000 km$^2$\,sr\,yr, more than the sum achieved with all other experiments.

Highlights of results include the clear evidence for a suppression of the flux above $4 \cdot 10^{19}$\,eV, observations of anisotropies in the arrival directions above $5.7 \cdot 10^{19}$\,eV suggesting a correlation to the nearby matter distribution, composition measurements favoring a change from a light to a heavier composition above $10^{19}$\,eV, a measurement of the $p$-Air and $pp$ inelastic cross-section at a cms energy of 57\,TeV, almost 10 times higher in energy than recent LHC data, and the most stringent upper limits on EeV photon and neutrino fluxes, strongly disfavoring an exotic particle physics origin for the highest energy cosmic rays. The Auger Observatory will continue running for at least 4 more years and upgrade plans are being discussed. 

\subsubsection{Telescope Array}

When AGASA and HiRes were nearing the end of operation, a collaboration consisting of key members from AGASA in Japan and the HiRes in US started to prepare for the construction of a large observatory, named Telescope Array (TA), in the Northern hemisphere. Like the Auger Observatory, TA combines a large area ground array, largely based on the AGASA design, with air fluorescence telescopes based on the HiRes system. TA is located in the central western desert of Utah, near the city of Delta, about 250\,km south west of Salt Lake City and covers with its 507 surface detector stations and 38 fluorescence telescopes a total area of about 730\,km$^2$. Data taking started early 2008 and because of this total number of events recorded is still much less then from the Auger Observatory. Nevertheless, good agreement within the systematic uncertainties is seen for the energy spectrum. Analyses of composition and anisotropies still suffer strongly from limited statistics; thus final statements need to wait for more data.

\section{Progress in answering the major questions}
\label{sec:progress}

The plan of this paper was to provide a description of the technical progress made towards answering the four major questions posed in the introduction.  While a historical article is not the place for a detailed progress report we shall now summarise briefly where the subject is with regard to answering each of these questions.

\subsection{Progress on understanding particle physics}

In the last two years data have become available from the LHC and it is satisfying to note that the models of hadronic interactions developed by cosmic ray physicists are found to be in good agreement with the new data \citep{Engel:2011fca,dEnterria:2011kga}.  Although the extrapolation from the Tevatron to LHC energies is over less than a decade, it suggests that the cosmic ray modellers are probably on the right track.  As already mentioned, the cross-section for $pp$-collisions can be inferred from shower data and a measurement has been reported based on Auger data at a cms energy of 57\,TeV.  This result is in good accord with the extrapolation from Tevatron and LHC measurements and suggests that the cross-section does not rise as fast with energy as predicted by some models \citep{Abreu-ppXsection-2012}.  However one is still far from fully understanding hadronic multi-particle production at the highest energies.

An incoming proton of $10^{17}$\,eV interacting with an air nucleus creates about the same energy in the nucleon-nucleon cms system as is reached in $pp$-collisions at the LHC.  It follows that the extrapolated input to hadronic models of such as inelasticity and multiplicity must be made for cms energies up to 30 times higher than those reached with man-made accelerators in the foreseeable future.  This reliance on extrapolation is seriously restricting as there is now clear evidence that some features of shower physics cannot be explained within the context of present models.  Specifically the primary-energy estimates made with models rather than from an empirical energy scale rooted in the fluorescence method, are found by both the TA and Auger groups to be too large by 20 - 30\,\%.  Further the Auger group finds, from various analyses, that the muon content of showers is substantially higher than predicted by any current model though, of course, the conclusion as to the magnitude of the muon deficiency with respect to each model does depend on the mass that is assumed.  
Thus there is more to learn about particle physics from the study of extensive air showers.

\subsection{Inferences from arrival direction distributions}

As already mentioned, the EAS-TOP team has observed the Compton-Getting effect at energies of $\sim 10^{14}$\,eV but there is no strong evidence for any other anisotropy although the intriguing alignment of the phase of the first harmonic in right ascension near 20 hours over a wide range of energies as first noted by the Cornell group is also seen in the lower energy data from the Auger Observatory \citep{Abreu-large-scale}.  There is still debate as to whether the distribution of arrival directions shows anisotropy at the highest energy.  Using the data set with the greatest statistics from the Pierre Auger Observatory, an association of events above $5 \cdot  10^{19}$\,eV with Active Galactic Nuclei has been reported at the level of 38\,\% \citep{Abraham-07e}. The probability of finding such a correlation, assuming isotropy, is 0.3\,\%. The measurements reported by the Telescope Array (TA) and HiRes groups are of lower statistical weight and, given the uncertainties in the absolute energy scales of $\sim 20$\,\%, it remains uncertain as to whether the Auger result is supported (TA) or refuted (HiRes).  

Thus it remains unclear as to what can be inferred from measurements of the arrival directions of high energy cosmic rays.

\subsection{The Energy Spectrum of Cosmic Rays}

In the region of the knee the shape of the energy spectrum has been very well-measured by a number of experiments, most notably by the KASCADE group, and the interpretation of their $N_\mu$ vs.\ $N_e$ plot has settled the question as to whether the knee feature is a manifestation of new hadronic physics or of acceleration or propagation effects.  The first possibility is now excluded. At higher energies, features of the energy spectrum are well-established with a flattening (generally known as the ankle) found at $4 \cdot 10^{18}$\,eV and a steepening at around $4 \cdot 10^{19}$\,eV \citep{Abbasi-08,Abraham-08c}.  

Thus the details of the cosmic ray energy spectrum are well-measured although the explanations for the features found remain under discussion and require an answer to the fourth question before further progress can be made.

\subsection{The mass composition of cosmic rays}

An important observable which is much harder to obtain than either the energy spectrum or the arrival direction distribution is the mass composition.  As already mentioned, significant progress has been made in the region of the knee and it is clear that there the mass is becoming heavier as the energy increases.  Just below $10^{17}$\,eV the KASCADE-Grande group, as already noted, have found evidence of what may be an iron-knee thus opening the possibility that there may be a transition to different sources at this energy.  Whether these sources are galactic or extra-galactic is subject to considerable controversy.

Although it is recognised that a parameter such as the depth of shower maximum, $X_{\rm max}$, is sensitive to the mass of the incoming primary particle, with showers initiated by heavy nuclei having smaller $X_{\rm max}$ than those initiated by protons, the differences are relatively small ($\sim 15$\,\%) and there are also fluctuations in shower development so that identifying the atomic mass, $A$, of a particular event, is presently not feasible.  To interpret the $X_{\rm max}$ measurements made by HiRes, Auger and TA it is necessary to compare the average $X_{\rm max}$ against predictions made with Monte Carlo simulations that use extrapolations of data from accelerator energies \citep{Abraham-xmax-10}.  Operation of fluorescence detectors demands clear moon-less nights resulting in an on-time that is less than 15\,\% and combined with the cuts necessary to obtain an unbiased sample this limits the statistical accuracy at the highest energies.  Selection effects are a problem that may not yet have been fully solved.

Using a range of models, developed before the first LHC data became available, the Auger measurements of $X_{\rm max}$ suggest that the mean mass of the primary particles increases with energy above about $3 \cdot 10^{18}$\,eV.  As with the arrival direction distributions, there is no consensus on this conclusion with the HiRes group asserting that the cosmic ray beam is proton-dominated at all energies above $\sim 10^{18}$\,eV.  Data from TA are statistically more limited and uncertainty as to the absolute energy scale again confuses the situation.  The question is of great importance as if the beam was dominated by protons at the highest energy then the feature at the ankle might be interpreted as evidence of pair-production as has been advocated forcefully by Berezinsky \citep{berezinsky-06b}, while the steepening at $4 \cdot 10^{19}$\,eV might be considered as evidence for the GZK-effect.  One would also have an important constraint on the environments in which acceleration can take place. For example a gamma-ray burst source would be proton-dominated if acceleration occurred in the jet but would reflect the composition of the surrounding inter-stellar medium if acceleration was at the termination shock. The role of GRBs as proton accelerators has been challenged by the IceCube neutrino observatory very recently \citep{Abbasi:2012tx}.

Knowledge of the mass is also required for reliable predictions of the flux of high-energy neutrinos with relatively fewer expected if the mass spectrum of ultra-high energy is dominated by heavy nuclei.  If Nature is kind enough to provide unequivocal evidence for a point source then we would surely be rather certain that the incoming primaries were protons and be able to resolve some of the astrophysical questions as well as those posed by the hadronic physics.  However this is for the future and we turn now to the projects that are in the pipeline.

\section{Future}
\label{sec:Future}

As discussed above, the situation at the upper end of the cosmic ray energy spectrum has changed considerably with the advent of new large scale observatories. No doubts anymore exist about the presence of a flux suppression above $\sim 5 \cdot 10^{19}$\,eV. However, is this the observation of the GZK effect which was predicted 45 years ago? From the experimental point of the view, the answer cannot be given, because the suppression could equally well be due to the limiting energy reached in nearby cosmic accelerators, just as discussed by Hillas in his seminal review [Hillas 1984]. In fact, the latter picture is supported by data from the Pierre Auger Observatory which suggest an increasingly heavier composition towards the end of the spectrum and seeing the suppression about 20\,\% lower in energy than expected for typical GZK scenarios. HiRes and TA, on the other hand find no significant change in their composition and their cut-off energy is in agreement with the GZK-expectation. Moreover, a directional correlation of ultra high-energy cosmic rays on a $3^\circ$ scale is hard to imagine for heavy primaries. Could this indicate weaker extragalactic magnetic fields than thought, or could it point to deficiencies of hadronic interaction models at the highest energies? These models must be employed to infer the elemental composition from EAS data.

Obviously, Nature does not seem ready to disclose the origin of the most energetic particles in the Universe yet. More work is needed and the main players in the field have intensified their co-operation sharing data and analysis strategies to better understand systematic uncertainties which, despite being small, appear to be quite relevant concerning conclusions to be drawn from the data. In parallel, experimental efforts are underway to increase the statistics more quickly and to further improve data quality. Most importantly muon detection capabilities, which are of key importance to understanding features of hadronic interactions at the highest energies, are being added. 

Understanding the origin of ultra high-energy cosmic rays demands high quality data in the $10^{19}$ to $10^{20}$\,eV energy range. While this is to be the major task of ground based experiments during the next years, finding the long awaited point sources of cosmic rays simply requires much larger exposures. Plans for space-based experiments exist as well as for further efforts on the ground.

\subsection{Going into Space}
In 1979 Linsley developed the idea to observe giant air showers from space \citep{Linsley-79}. The advantages were obvious, as a fluorescence camera looking downwards from space could survey huge areas at ground simultaneously with only one atmospheric thickness between the light source and the sensor. The major challenge was the faint light because of the distance to the shower and the optical imaging required for geometrical reconstruction and $X_{\rm max}$ observations. The initial project was called SOCRAS (Satellite Observatory of Cosmic Ray Showers). Y Takahashi took up this idea in the 1990s and developed it further. MASS (Maximum-energy Auger air Shower Satellite) was presented at the 1995 ICRC in Rome \citep{Takahashi-95} and used Fresnel optics to enlarge the field of view to $\pm 30^\circ$ yielding an observational area of 100\,000 km$^2$. The next years were followed by ups and downs, most importantly by the tragic space shuttle Columbia disaster on 1st February 2003 which put all plans on hold. The possibility to mount the Extreme Universe Space Observatory (EUSO), originally submitted as an ESA proposal, on the exposure facility of the Japanese Experiment Module (JEM) of the ISS offered a new window of opportunity and JEM-EUSO is planned to be launched in 2015-2016. Further projects for measuring ultra high-energy cosmic rays from space are TUS to be launched in 2012 as a pathfinder on the Russian satellite Lomonosov, KLYPVE an MSU/ROSCOSMOS mission involving a 4\,m Diameter lens with on-orbit assembly by astronauts, and on the long term possibly S-EUSO. The realisation of the missions involves some uncertainty and it is clear that the energy and mass resolutions for cosmic rays will be much worse than that achieved with ground based observations. Their prime goal is to collect event statistics at the highest energies to detect the long-searched point sources of ultra high-energy cosmic rays.

\subsection{Old Technologies Revisited}

\subsubsection{Radio and Microwave Observations}

The successful detection of of air-Cherenkov radiation by Galbraith and Jelley \citep{Galbraith-53} discussed in Sec.\,\ref{sec:PMT} did at that time not seem to offer great potential for EAS detection because of being concentrated in the forward direction with a lateral spread similar to that of secondary particles and because of being limited to clear moonless nights reducing the duty cycle of observations to about 10\,\%. However, this observation triggered Jelley to consider whether the Cherenkov emission mechanism that gives rise to optical emission with a $\nu \Delta \nu$ spectrum, might also radiate in the microwave region of the spectrum and possibly be detectable with sensitive receivers \citep{Jelley-58}. He was not very optimistic, however. In 1962 G Askaryan published a short paper in which he suggested that the particle cascade resulting from the interaction of a high energy particle in a dense medium would not be electrically neutral since the resulting positrons could decay in flight. Also, the cascade would accumulate delta-rays and Compton scattered electrons \citep{Askaryan:1962ts}. Cherenkov radio emission could then occur at longer wavelengths where there would be coherent emission from the net negative charge. Askaryan also pointed out that geomagnetic effects might  contribute to separation of the charged components in the shower and this dipole might provide an additional emission mechanism. The emission would be coherent when the dimensions of the shower-emitting region become comparable with the wavelength, offering a huge gain for the signals to be observed.
%
To preserve coherence, the radiating particles all have to be at the same distance from the detecting antenna, to an accuracy of a fraction of a wavelength. Since the longitudinal dispersion of the shower particles is effectively the shower disk thickness of less than 3\,m, the coherence condition requires that the wavelength of observation be greater than the physical dimensions of the emitting region. At a frequency of 75\,MHz the corresponding wavelength is 4\,m, so the observational wavelength needs to be greater than this value for detection of a coherent signal.
When Porter heard about this paper, he contacted Jelley and they conducted a simple experiment at the Jodrell Bank Radio Observatory in 1964. The historical details which resulted in the observation of 11 events measured in coincidence with Geiger counters are well described at first-hand by T Weekes \citep{weekes01} and D Fegan \citep{Fegan-12}. The observation resulted in a flurry of activities but none of these experiments gave conclusive results on the nature of the emission.
A most comprehensive review article by H Allan was published in 1971 \citep{Allan-71}. By this time it had been recognised that the rather steep fall-off of the radio signals at all frequencies made it difficult to envisage the construction of a large shower array using the radio technique alone.  The spacing of the detectors would need to be closer than 1000\,m, which has consequences for land access and costs.  Work continued with the aim of using the technique to study the longitudinal development of showers and hence gaining information about the mass composition.  However a further difficulty was recognised in that the magnitude of the atmospheric electric field, the variation of which was hard to monitor on short timescales in the 1970s, has a significant effect on the magnitude of the radio signals as summarised by Watson \citep{Watson-75}. Moreover, particle detectors and fluorescence observations seemed to offer large and more easily accessible potential for cosmic ray detection.

With the advent of digital logic hardware, powerful low-cost computing, the ability to perform Monte Carlo simulations and above all the needs to considerably extend the aperture of ultra high-energy experiments interests in radio observations revived explosively a decade ago. D Saltzberg and P Gorham verified in beam measurements at SLAC the existence of the Askaryan effect in dense media \citep{saltzberg01}, H Falcke and Gorham in 2003 revived the possibilities of measuring ultra-high energy cosmic rays and neutrinos with radio techniques \citep{falcke-03}, and Falcke and Kampert developed the idea of the LOPES experiment to test the potential of the radio technique with state of the art electronics at the KASCADE air shower experiment which convincingly demonstrated the detection and imaging of atmospheric radio flashes from cosmic ray air showers with the geo-synchrotron effect as the main underlying mechanism \citep{Falcke-05}. Similar observations were made independently at the CODALEMA experiment in France. With the Netherlands entering the field, scientists from all three countries have joined forces to construct the Auger Engineering Radio Array (AERA) at the Pierre Auger Observatory. The goal of the ongoing activities is to verify the practicality of the radio technique for a giant future observatory and to explore the performance for energy and mass measurements. Moreover, radio observations by balloon-borne experiments offer the possibility of surveying huge areas with only a few antennas. In this case, the reflected radio beam off the surface is being detected. The feasibility of such a concept has been demonstrated very recently by the ANITA experiment flown over Antarctic ice \citep{Hoover:2010id}.

Recently, again triggered by Gorham et al.\ \citep{Gorham:2008fa}, the possible detection of microwave radiation from cosmic ray extensive air showers became another revived topic of ongoing experimental activities. The continuum radiation in the microwave range is expected to be caused by free-electron collisions with neutrals in the tenuous plasma left after the passage of the shower. Again the process seems to be confirmed by accelerator experiments, but the proof of emission from EAS remains to be shown. Unlike the geo-synchrotron and Cherenkov emission, the microwave emission should occur isotropically which would make it an extremely powerful experimental tool, if confirmed.

\subsubsection{Radar Observations}
\label{sec:radar}

The possibility of detecting very large air showers by reflecting a radar beam from the ionisation column that they create in air was first pointed out by Blackett and Lovell \citep{Blackett-41b} who calculated that the showers should produce enough ionisation to give a detectable reflection of a suitable radio signal.  The radio signal that they had in mind was the type being used at that time in radar detectors.  Indeed according to Lovell \citep{Lovell-93} it was the observation of transient signals on radar screens, made by himself and J G Wilson in September 1939, that prompted the idea.  Of course the radar technique was secret at this time.

Soon after the publication of the 1941 paper it was pointed out by T L Eckersley in a letter to Blackett that the estimate of the recombination time of the electrons in the shower plasma was too long by a factor of $\sim 10^6$.  Because of other priorities this letter was overlooked until the end of the WW\,II and was not studied by Blackett and Lovell before Lovell, at Blackett's insistence, had taken radar equipment to Jodrell Bank to search for showers.  By the time the implications of the short recombination time were appreciated, evidence that most of the reflections were from the ionisation trails that he and Wilson had observed were associated with the ionisation left by meteors.  Lovell notes that in the scientific case made for building the 250 foot steerable Jodrell Bank telescope the detection of radar echoes from showers was included.

Further efforts to implement this idea have been made from time to time.  Suga \citep{Suga-62} and Matano et al.\ \citep{Matano-68} were the first to revive the idea using a Loran station of 100 kW at 1.85 MHz near Tokyo.  However the area proved to be noisy and no events were detected.  A further theoretical study was made by Gorham \citep{gorham01} and this led to a further but again unsuccessful attempt in Japan by Iyono et al.\ \citep{Iyono-2003}.  More recently, Wahl, Chau and Bellido \citep{Wahl-07} have used the Jicamarca radio observatory (JRO) in Peru (50 MHz and 2 MW) and have detected some anomolous reflections in a search of JRO meteor data.
A further attempt is being mounted at the Telescope Array and KASCADE sites.

\section{Concluding Remarks}
\label{sec:Conclusions}

In this year, 2012, the centennial of the discovery of cosmic rays will be celebrated all around the globe. The enormous progress that has been made during this period is directly linked to the invention of new experimental tools and instrumentation and could not have been made without the ideas and skills of some ingenious pioneers. Almost no nuclear and particle physics experiment could be done without making use of the coincidence technique but also triggering on rare events, and the construction of calorimeters, as other concepts, have been pioneered in cosmic ray experiments. To remain focussed on cosmic ray and air shower physics, we have omitted in this review the discoveries of new particles made by cosmic ray observations, including the positron, muons, pions, kaons, hyperons, and likely also charmed particles. This part of the history is discussed in \citep{walter-epjh}.

The cosmic energy spectrum has been measured in great detail over more than 32 decades in flux, making this observable unique in Nature. The spectrum initially thought to follow a pure power law distribution has exhibited more and more structure, starting with the discovery of the ``knee'' at about $4 \cdot 10^{15}$\,eV by Khristiansen's group at Moscow State University in 1959, followed by the observation of the ``ankle'', first hinted at by Linsley \citep{Linsley-63a}, at Haverah Park, Akeno, and Fly's Eye in 1991 and the suppression at the GZK threshold in 2008 by the HiRes and Auger observatories. Very recently, a second knee caused by the heavy cosmic ray component has been reported by KASCADE-Grande and it is not unlikely that even more departures from a simple power-law distribution will be exhibited providing important clues about the origin of cosmic rays. Also, great detail about the primary mass could be extracted from the data with remarkable changes seen in the composition coinciding with the the position of the structures in the energy spectrum \citep{Kampert-12}. The sky in cosmic rays is surprisingly isotropic up to the highest energies and is challenging our understanding of both cosmic ray propagation within the galactic and intergalactic environments and about their sources. Only at the highest energies are departures from isotropy are seen, but data suffer still from statistics.

Particles at the upper end of the spectrum have such breath-taking energies, a hundred million times above that provided by the LHC accelerator, that the questions about how cosmic accelerators can boost particles to these energies, and about what is the nature of the particles themselves, are still open and of prime interest. The mystery of cosmic rays is nowadays tackled - and is perhaps going to be solved - by an interplay of sophisticated detectors for high-energy $\gamma$-rays, charged cosmic rays and neutrinos. Moreover, plans for next generation experiments are being worked out and it is now realized that the true high-energy frontier in Nature provides unique opportunities to test particle and fundamental physics, such as of space-time, at its extreme. Further surprises by future cosmic ray observations are almost guaranteed.

\section*{Acknowledgments}
We gratefully acknowledge stimulating discussions with our colleagues in the Auger and KASCADE-Grande collaborations. We also thank Luisa Bonolis, Antonella Castellina, Bruce Dawson, Piera Ghia, Antoine Letessier-Selvon, Marco Segala, and Mike Walter for their generous help to access some of the original key papers distributed at various archives around the world and Dave Fegan for critical reading and many constructive suggestions.
KHK also acknowledges financial support by the German Ministry for Research and Education (BMBF) and by the Helmholtz Alliance for Astroparticle Physics, and AAW acknowledges financial support from the UK Scientific, Technology and Facilities Council and Leverhulme Foundation.

\bibliographystyle{plainnat}
\bibliography{CR-Showers-EPJ}

\end{document}